\documentclass{article}

\usepackage{arxiv}

\usepackage[utf8]{inputenc} 
\usepackage[T1]{fontenc}    
\usepackage{hyperref}       
\usepackage{url}            
\usepackage{booktabs}       
\usepackage{amsfonts}       
\usepackage{nicefrac}       
\usepackage{microtype}      
\usepackage{lipsum}
\usepackage{natbib}
\usepackage{graphicx}
\usepackage{float}
\usepackage{algpseudocode} 
\usepackage{placeins}
\usepackage{siunitx}
\usepackage{graphicx}
\usepackage{wrapfig}
\usepackage{multirow}

\usepackage{enumitem}   
\graphicspath{ {./images/} }

\title{Implicit Neural Models for Direct Multi-Shell Diffusion MRI Signal Estimation
}

\author{
Sanna~Persson\thanks{Equal contribution.}~\thanks{Corresponding author: \texttt{sannape@kth.se}} \\
Biomedical Engineering and Health Systems \\
KTH Royal Institute of Technology \\
Stockholm, Sweden \\
\texttt{sannape@kth.se} \\
\And
Fabian~Leander Sinzinger\footnotemark[1] \\
Biomedical Engineering and Health Systems \\
KTH Royal Institute of Technology \\
Stockholm, Sweden \\
\texttt{fabiansi@kth.se} \\
\And
Christoffer Olsson \\
Biomedical Engineering and Health Systems \\
KTH Royal Institute of Technology \\
Stockholm, Sweden \\
\texttt{chrol@kth.se} \\
\And
Rodrigo Moreno \\
Biomedical Engineering and Health Systems \\
KTH Royal Institute of Technology \\
Stockholm, Sweden \\
\texttt{rodmore@kth.se} \\
}

\begin{document}
\maketitle

\begin{abstract}



Diffusion magnetic resonance imaging is a noninvasive technique that is sensitive to the motion of water molecules in human tissue. The image signal is commonly described in q-space; however, a complete coverage of this space is often infeasible due to hardware and acquisition time limitations.
In this work, we propose to represent the entire diffusion signal using an implicit neural representation network with wavelet activation functions. The signal space is encoded using Fourier features and spherical harmonics basis functions for a continuous output distribution of the raw diffusion data, enabling both super-resolution and q-space interpolation.
Additionally, we introduce a pretraining strategy using a group template, enabling faster per-subject optimisation when combined with nonlinear registrations of subjects into template space. We tested the method with 10 subjects from the Human Connectome Project and 26 subjects from in-house data (multidimensional diffusion MRI from a Parkinson's disease cohort). 
The results show that the proposed method yields high-fidelity reconstructions with significantly reduced memory requirements compared to raw diffusion-weighted images. The pretraining scheme showed good generalisability properties on data acquired with very different protocols. We further assess the method on downstream metrics relevant to diffusion MRI data from the learned representations. 

\textbf{Keywords:} Multi-Shell Diffusion Magnetic Resonance Imaging, 
Implicit Neural Representations, 
PQ-Space, 
Spherical Harmonics, 
Representation Learning, 
Brain Image Processing, 
Signal Interpolation,


\end{abstract}

\section{Introduction}\label{sec:intro}

Diffusion‐weighted magnetic resonance imaging (DWI) enables non-invasive characterisation of tissue microstructure by sensitising the MR signal to the random Brownian motion of water molecules \cite{basser_mr_1994}. By modelling the diffusion process, techniques such as diffusion tensor imaging (DTI) have revolutionised the study of brain connectivity and are increasingly used in other fields such as oncology \cite{le_bihan_what_2019}, musculoskeletal imaging \cite{wang_anisotropic_2025} and cardiovascular imaging 
to probe microscopic tissue properties \cite{jelescu_neurite_2022}.  

DWI comprises a set of 3D images acquired at different b-values and b-vectors, two parameters that span the so-called q-space. Specifically for a point $\textbf{q}$ in q-space, the b-value, $b$, and b-vector, $\mathbf{\hat{q}}$, correspond to its the magnitude and orientation of $\textbf{q}$, respectively. 
Thus, DWI is inherently a 6D imaging modality with three spatial coordinates and three q-space coordinates \cite{wedeen_mapping_2005}. The b-values and b-vectors are selected in the MRI scanner by manipulating the machine's magnetic field gradients. DWI acquisition is usually restricted to a very reduced number of sampling points of the q-space due to the fact that the acquisition time linearly increases with this number. These sampling points are usually distributed over multiple spherical shells with different radii (multi-shell DWI). 

Early single-shell protocols captured diffusion along a limited number of b-vectors and a single b-value, constraining the biological specificity of downstream models. Multi-shell acquisitions, which sample multiple b-vectors at different b-values, enable a better estimation of signal representations such as fibre orientation distributions, which in turn have an impact on the quality of tractography used in structural connectivity analyses and microstructure estimations. 

Methods to increase spatial and q-space resolution are promising in addressing the issues of current DWI data acquisition protocols. In the last few years, different data-driven methodologies have been proposed for image super-resolution, in which higher-resolution images are estimated from lower-resolution images while restoring missing details. In particular, implicit neural representations (INRs) have been shown to be one of the most successful methods for image super-resolution. An additional advantage of INRs is that it is able to compress data, which is very useful for DWI data.

In this paper, we introduce a method for representing the raw DWI signal with INRs. We show that pretraining the INRs with a DWI template is effective to drastically reduce the training time for unseen subjects. We further demonstrate capabilities for sparsely sampled data completion. 


\section{Related work}
Our work builds upon research in both DWI signal prediction and INR models. In this section, we provide an overview of previous approaches to DWI signal prediction.

\textbf{Spatial super-resolution}: Traditionally, super-resolution has been targeted using interpolation. Deep learning-based methods have been shown to be superior for this task, including DWI data (e.g. \cite{guo_semi-supervised_2023,chatterjee_shuffleunet_2021}). 

\textbf{Intrashell interpolation}:
A straightforward approach for DWI interpolation in a specific shell is to fit the acquired data using spherical harmonics and sample the unseen data at any orientation \cite{descoteaux_regularized_2007}.
Alternatively, \cite{jumakulyyev_combining_2023} proposed an in-painting strategy that involves solving partial differential equations, though this approach can be computationally expensive. In \cite{ren_q-space_2021}, a neural network was used to estimate DWI from structural images. These methods are restricted to the b-values used during training. Convolutional neural networks (CNNs) and visual transformers have also been proposed for intra-shell interpolation \cite{elaldi_equivariant_2021,elaldi_e3_2024}. 
Beyond DWI signal prediction, there exist methods to predict downstream parameters such as mean diffusivity and fractional anisotropy \cite{aja-fernandez_validation_2023,golkov_q-space_2016,gu_generating_2019}. The main limitation with these methods is that, unlike INRs, they are restricted to interpolations at a discrete grid. %

\textbf{Single-shell to multi-shell prediction}  Going from single-shell to multi-shell DWI is an interesting avenue for downstream DWI representations that require multiple shells. One can design a physical model for continuous predictions, as done for example in intravoxel incoherent motion (IVIM) \cite{le_bihan_what_2019}, diffusion kurtosis imaging \cite{henriques_toward_2021}, or more complex models such as SHORE \cite{ozarslan_mean_2013}. Data-driven approaches are, however, appealing as they are general-purpose and do not rely on any model. For example, CNNs have also been used to
predict DWI from other shells \cite{jha_single-shell_2022}, although it is limited to the b-values used for training.

\textbf{Implicit neural representations for DWI:} One of the main advantages of INR is that they can efficiently tackle all the previously discussed problems simultaneously. The use of INRs for DWI data is relatively new. One of the first studies that used INRs for DWI data found that INRs are effective and efficient in compressing DWI data \cite{mancini_lossy_2022}. INRs have been used for super-resolution DWI in prostate imaging \cite{gundogdu_self-supervised_2024}. Moreover, two studies have shown the promise of INRs to predict fODF fields from mono-shell \cite{consagra_neural_2024} and multi-shell DWI data \cite{hendriks_implicit_2025}, respectively. An issue with these two methods is that predicting fODFs instead of the DWI signal restricts the application to tractography. To our knowledge, only NeSH \cite{hendriks_neural_2023} has targeted the prediction of the DWI signal using INRs. The authors show that the method performs better for intrashell interpolation and super-resolution than spherical harmonics interpolation and cubic interpolation, respectively. 
Unlike NeSH, our method can also be used for continuous multi-shell DWI signal prediction. 

\textbf{Contributions:}
One of the limitations with training INRs is that the training is subject-specific, and for each data point, a new model fit is necessary. There has been previous work to develop meta-learning methods \cite{lee_meta-learning_2021,paolis_fast_2024,vyas_learning_2024} to mitigate this limitation and improve adaptation to unseen data. In the present work, we explore a related approach by pretraining our network on a group template in the same signal space. While commonly used group templates exist for structural MRI \cite{fonov_unbiased_2009,fonov_unbiased_2011}, and even derived fibre distribution templates have been proposed \cite{lv_building_2023}, there is, to the best of our knowledge, no widely adopted multi-shell DWI template.
Consequently, we develop a group-averaged DWI template from HCP subjects to pretrain INRs and facilitate faster subject-specific optimisation. Moreover, as mentioned, our method can also perform interpolation across shells. We further evaluate three INR architectures and settings and quantitatively assess the performance at downstream parameter estimation tasks.  


\section{Methods}

\subsection{Signal Representation as a Continuous Function} 


Diffusion-weighted imaging (DWI) is commonly described as a continuous, voxel-wise signal defined on a spatial domain $P \subset \mathbb{R}^3$. In practice, however, images are discretised on a regular voxel grid, restricting the signal domain to $\bar{P} \subset \mathbb{Z}^3$. Consequently, the image can be expressed as a function $I: \hat{P} \to Y$, where $\mathbf{y} \in Y$ denotes the signal value, which may be scalar, vector-valued (e.g., multi-channel), or equipped with additional structure, such as multi-shell q-space signals.

While this discrete formulation is well-suited for storing image data in multidimensional arrays, it does not always meet the practical requirements of brain diffusion pipelines, where a continuous, structure-preserving representation is often preferable (e.g., for tractography or nonlinear registration). We therefore propose to model the multi-shell DWI signal as a scalar-valued function defined over a structured product space:
\[
I: P \times B \times Q \to \mathbb{R},
\]
where $P \subset \mathbb{R}^3$ denotes spatial coordinates, $B \subset \mathbb{R}_+$ represents the b-value domain, and $Q \subset \mathbb{S}^2$ corresponds to the angular domain of diffusion directions (i.e., the unit sphere). A coordinate in this product space is denoted by $\mathbf{x} = [x, y, z, b, \alpha, \beta]$. Compared to NeSH \cite{hendriks_neural_2023}, inputting $b$ allows for continuous multi-shell DWI signal prediction instead of single-shell.

\begin{figure}
    \centering
    \includegraphics[width=1.\linewidth]{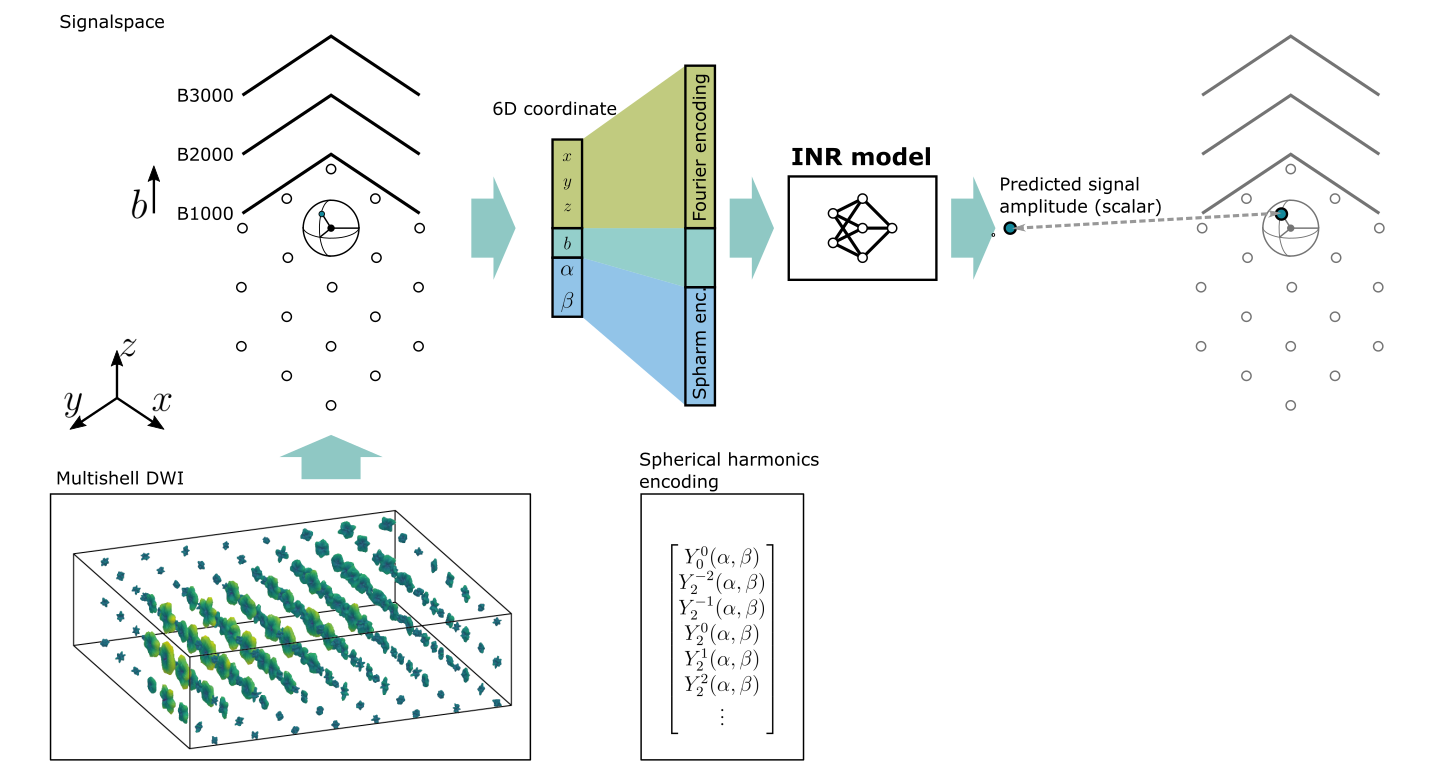}
    \caption{Overview of the proposed representation for learning the DWI signal using an implicit neural representation (INR). The multi-shell DWI is represented as pairs of signal-space coordinates and corresponding scalar amplitude values. The spherical q-space signal-space coordinates are encoded via spherical harmonics function evaluations (bottom, centre).}
    \label{fig:overview}
\end{figure}

\subsection{Input Encoding} 



In previous work on INRs, appropriate encodings of the input before the INR network have been shown to significantly improve representation learning (e.g., \cite{tancik_fourier_2020}).
Following conventional practice, we encode the spatial and b-value coordinates $(x, y, z, b)$ via Gaussian random feature encoding. For the q-space coordinates $\mathbf{p} = (\alpha, \beta) \in \mathbb{S}^2$, we propose to encode them by evaluating the real spherical harmonic basis functions up to degree $\ell_{\text{max}}$ ($\ell_{\text{max}} = 8$ in our experiments) at the respective position $\mathbf{p}$ (see Figure~\ref{fig:overview}, box b). 
Moreover, we restrict the basis to even degrees only ($\ell = 0, 2, \dots$), effectively enforcing antipodal symmetry of the learned representation. In other words, under the proposed spherical encoding, a point $\mathbf{p}$ and its antipode $-\mathbf{p}$ are mapped to the same feature vector. This encoding constrains the model to only representing functions consistent with the common DWI assumption that q-space signals exhibit antipodal symmetry.

\subsection{Implicit Neural Representation} \label{sec:inr_models}

We implement three different INR architectures for training the DWI template and evaluate their performance on learning the representation of the full diffusion table. The input to the model is the encoding of the spatial and q-space position of each data point from the previous section.   

\paragraph{Base ReLU model}
The base model consists of a three-layer perceptron with leaky ReLU activations, incorporating layer normalisation and skip connections. 

\paragraph{SIREN model}
To better capture high-frequency content, the activation function $\mathbf{h}$ is replaced by a sinusoidal function and the SIREN initialisation strategy is adopted \citep{sitzmann_implicit_2020}:
\begin{equation}
  \mathbf{h}_{i+1} \;=\;
  \sin\!\bigl( \omega_0\,W_{i}\,\mathbf{h}_{i} + \mathbf{b}_{i} \bigr),
  \qquad \text{with } \omega_0 = 30 \; \text{as in: }  \cite{sitzmann_implicit_2020}.
\end{equation}
$W_i$ and $\mathbf{b}_i$ are the weight matrix and biases for layer $i$, respectively. For layers $i\!>\!0$ the entries of $W_{i}$ are sampled from the uniform distribution
$\mathcal{U}\!\bigl[-\sqrt{6/n_{\text{in}}}\!/\omega,\,\sqrt{6/n_{\text{in}}}\!/\omega\bigr]$,
whereas the first layer employs
$\mathcal{U}\!\bigl[-1/n_{\text{in}},\,1/n_{\text{in}}\bigr]$.
This yields stable gradients while preserving a broad spectral response.

\paragraph{\textsc{Wire} model}
This model \citep{saragadam_wire_2023} generalises the SIREN approach with learned frequencies. Our implementation employs a Gabor-like activation with frequency and scale components dynamically generated for each hidden dimension through distinct linear transformations of the input activation: 
\begin{equation}
  \mathbf{h}_{i+1} = \cos \ \!\bigl( \omega_{h} (W_{\text{f},i}\mathbf{h}_{i} + \mathbf{b}_{\text{f},i}) \bigr) \odot \exp\ \!\bigl( -(\sigma_{h} (W_{\text{s},i}\mathbf{h}_{i} + \mathbf{b}_{\text{s},i}))^2 \bigr).
\end{equation}
Here, $W_{\text{f},i}, \mathbf{b}_{\text{f},i}$ and $W_{\text{s},i}, \mathbf{b}_{\text{s},i}$ are the weights and biases for the linear layers that output frequency and scale-related terms, respectively. The factors $\omega_{h}$ and $\sigma_{h}$ are fixed scalar hyperparameters, and $\odot$ denotes the Hadamard product. The weights for these transforming layers ($W_{\text{f},i}$ and $W_{\text{s},i}$) are initialised based on the SIREN scheme from a uniform distribution. 


\subsection{DWI Group Template for Pretraining} \label{sec:pretrain}



\begin{figure}[H]
    \centering
    \includegraphics[width=1.\linewidth]{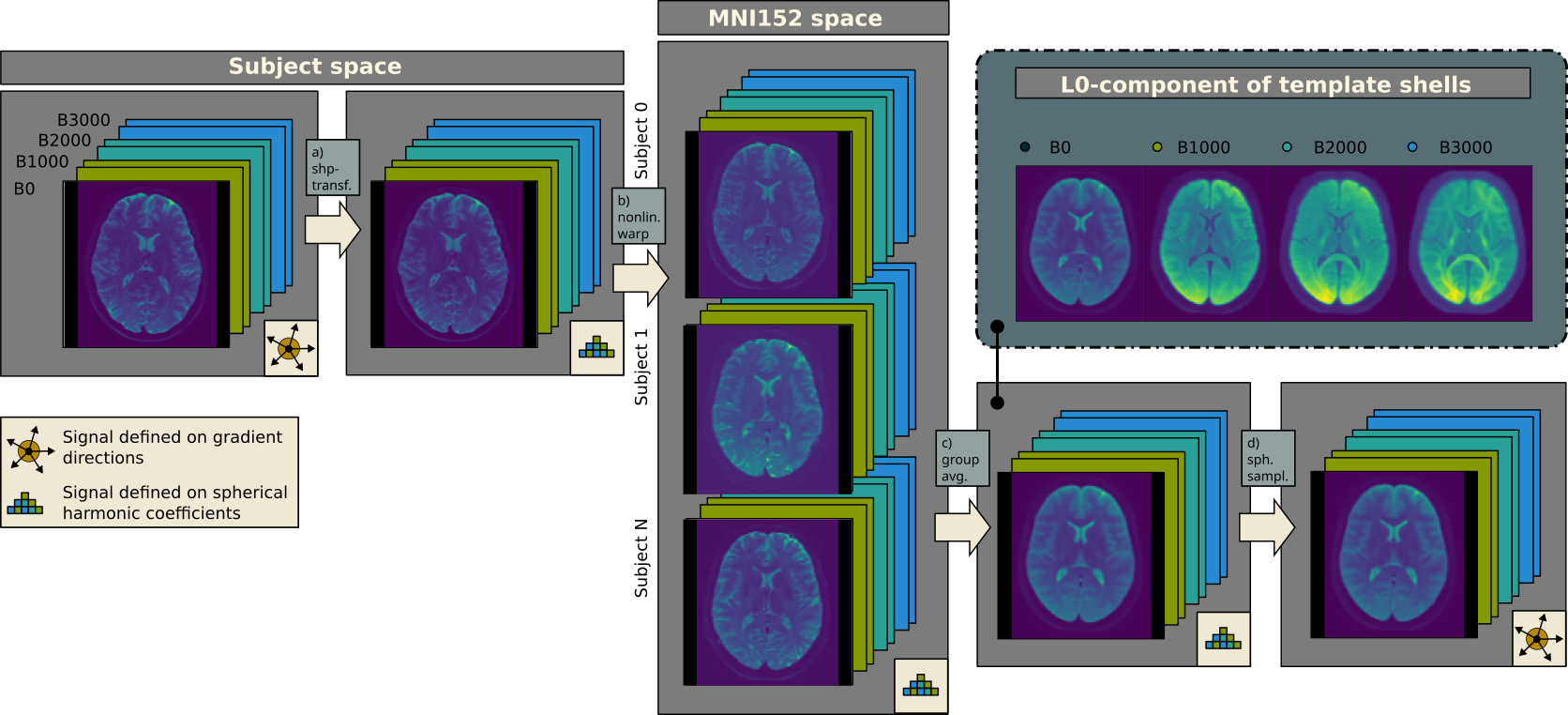}
    \caption{a) First, the DWI signal, defined at multi-shell gradient directions, is transformed into the spherical harmonics domain by performing a linear least squares fit of the harmonic coefficients per shell. b) Then, individual shells in spherical harmonics form are affine-transformed and subsequently nonlinearly warped to MNI152 space. The affine and nonlinear deformations were obtained from the official HCP preprocessing pipeline \cite{glasser_minimal_2013}. c) Next, the spherical harmonics coefficients are averaged across all 50 subjects. d) Finally, a DWI template in spatial form is reconstructed by sampling the spherical harmonics representation at discrete gradient directions.}
    \label{fig:templ}
\end{figure}

The aforementioned INR limitation, requiring subject-specific training, poses demands on computational cost and training time. In order to reduce the training time, we propose to train an INR on a DWI template that is later fine-tuned on the specific subject. Fig. \ref{fig:templ} shows the template generation process. First, we model the diffusion signal at every shell with spherical harmonics (SH) in subject space. The advantage of using SH instead of b-vectors is that b-vectors can vary from subject to subject \cite{andersson_integrated_2016}, which is not an issue with SH modelling. In the second step, we non-linearly transform the images into the MNI152 space using MRtrix3 \cite{tournier_mrtrix3_2019, raffelt_reorientation_2012}.\footnote{Specifics about the nonlinear deformation process are detailed in the appendix.} Finally, the SH representations in MNI152 are averaged and sampled back to b-vector representations.

In particular, we used 50 subjects from the Human Connectome Project (HCP) to create the template due to its high angular and spatial resolution at higher b-values. While we do not have access to lower b-values for generating the template, we will show that datasets with lower b-values also benefit from this template. Notice that \cite{lv_building_2023} proposed a similar methodology for creating templates, but for creating a model of fiber orientation distributions (FODs). By targeting the diffusion signal instead of FODs, we expand the scope of DWI applications beyond tractography (e.g., microstructure estimations).



\section{Experiments}
We focus our experiments on evaluating the brain template's effectiveness, q-space interpolation and downstream DWI metrics' quality. Through the experiments, we aim to give insight into the following research questions. 

\textbf{RQ1:}Does pretraining on a DWI template accelerate learning of subject-specific and template-registered signal representations?\\
\textbf{RQ2:} How sensitive is the INR model’s reconstruction quality to varying the number of training samples (b-vectors)?\\
\textbf{RQ3:} How effectively can the INR model interpolate diffusion MRI signals across different b-values?\\
\textbf{RQ4:} Can downstream metrics from diffusion MRI, specifically mean diffusivity and fractional anisotropy be reconstructed?

\subsection{Datasets}\label{data}

\paragraph{Human Connectome Project}
The Human Connectome Project (HCP) Young Adult cohort offers one of the largest publicly available collections of high-quality multi-shell diffusion MRI (dMRI) sampling 1200 healthy adults aged 22-35 years \citep{van_essen_human_2012}. 
Each acquisition consists of three interleaved b-value shells (b = 1000, 2000 and 3000 s/mm$^{2}$) with 90 uniformly distributed diffusion encoding directions per shell and six interleaved b = 0 images, yielding 288 volumes per scanning session across 282 unique points in q-space \citep{sotiropoulos_advances_2013}. 
For the fine-tuning from the template, the subjects are sub-sampled to 2x2x2 $\unit{mm}^3$ voxel size and non-linearly registered to the template MNI152 space. For all experiments we select a random subset of the minimally pre-processed volumes from the S1200 subjects release, detailed in the Supplemental material.

\paragraph{Multi-Dimensional Diffusion Data on Parkinson's Disease Cohort}
Multi-dimensional diffusion MRI (MD-dMRI) is an advanced technique in which 2nd-order tensors (b-tensors) are used instead of b-vectors \cite{topgaard_multidimensional_2017}. We have collected MD-dMRI data from 29 adults (17 healthy controls, 12 individuals with Parkinson’s disease) with the methodology described in \cite{olsson_effects_2025}.
The demographic and clinical profile of all participants is provided in the Supplementary material.
The MD-dMRI images were acquired with the process described by Topgaard et al.\cite{topgaard_multidimensional_2017} 
In particular, the linear b-tensors are conceptually equivalent to b-vectors of DWI. For this reason, we only used the data acquired with linear b-tensors in this paper.
The linear acquisition has 44 linear b-tensors spread over the b-values (b=0, 100, 700, 1400, 2000 $\unit{s}$/$\unit{mm}^{2}$), as adopted from \cite{szczepankiewicz_linear_2019}. 
Ethical approval was obtained and all subjects signed an informed consent. 
The data was anonymised and skull-stripped. 
The subjects were non-linearly registered to template space using ANTs \cite{tustison_antsx_2021,tustison_antsx_2024}. Visual inspection revealed that 3 subjects failed the registration and were therefore removed from the experiments, leaving us with 26 subjects. In the following, we refer to this dataset as ParkMR.



\subsection{Experimental setup}
Each model is trained with the mean squared error loss function, Adam optimizer and cosine annealing for at most 64 epochs. For retraining, we limit training time to 16 epochs unless otherwise specified. We normalize the DWI signal to be between 0 and 1 and report results on normalized images since diffusion MRI is not a quantitative modality. The training configuration for each experiment is found in the Supplemental material. For evaluation we report mean squared error (MSE), mean absolute error (MAE), relative error (RE) with respect to MAE, peak signal to noise ratio (PSNR), structural similarity score and the LPIPS perceptual metric using AlexNet.

\subsection{Comparison among different architectures}

We conducted an initial experiment to select one of the three INR candidate models from Section~\ref{sec:inr_models} for further exploration. Table~\ref{tab:overall_metrics} compares their performance when trained to reconstruct the DWI group template introduced in Section~\ref{sec:pretrain}. The DWI template was used for this experiment as it aggregates diffusion signals from 50 HCP subjects (distinct from the 10 used for further testing), thus minimising subject-specific bias. As shown in Table~\ref{tab:overall_metrics}, \textsc{Wire} consistently achieves the best overall results. Therefore, all subsequent experiments utilise the \textsc{Wire} architecture.

\begin{table}[H]
  \centering
  \small
  \caption{Overall reconstruction performance (averaged across all voxels and shells) for the DWI template.
           Lower is better for MSE/MAE/RE/LPIPS, higher is better for PSNR/SSIM.}
  \label{tab:overall_metrics}
  \begin{tabular}{lcccccc}
    \toprule
    Method & MSE $(\times10^{-4})$ $\downarrow$ & MAE $\downarrow$ & RE (\%) $\downarrow$ &
            PSNR (dB) $\uparrow$ & SSIM $\uparrow$ & LPIPS $\downarrow$\\
    \midrule
  Leaky  & 2.52 & 0.0113 & 7.05 & 32.37 & 0.972 & 0.0164 \\
    SIREN  & 1.81  & 0.0100 & 7.35  & 33.75 & 0.978 & 0.0120 \\
    \textsc{Wire} & \textbf{0.73} & \textbf{0.0063} & \textbf{4.43} &
                   \textbf{37.74} & \textbf{0.991} & \textbf{0.0041} \\
    \bottomrule
  \end{tabular}
\end{table}

\subsection{Pretraining on group-averaged DWI template}


In Tables~\ref{tab:overall_variants} and~\ref{tab:overall_variants_ParkMR}, we report the performance of \textsc{Wire} across the HCP and ParkMR datasets. We trained four different training setups per subject. Two setups train in the original subject space: one training from scratch ("Unregistered") and one using pretraining ("Pretrained Unreg"), and two trained on non-linearly registered subjects to our DWI template with and without pretraining ("Registered" and "Pretrained Reg"). As shown, training on the registered images consistently led to better results in the evaluations we conducted.

We furthermore observe that pretraining on the DWI template is effective: models fine-tuned for only 25\% of the training time (16 epochs) reached a very similar performance as models trained from scratch for 64 epochs. This speed-up can be used for either faster training or improving performance. This effect was clearly visible for registered images, but less pronounced for unregistered ones. This was expected as registration aligns the training coordinates with those seen during pretraining. While pretraining was expected to be less useful for unregistered data, we found that pretraining also reduces training time in this case, although to a lesser extent.

 It should be noted that in HCP, registration involves downsampling to 2\,mm MNI space, reducing the number of training points. In ParkMR, it leads to upsampling and increases the number of training points. Despite this, registration still improves performance in both cases. We also observe a slight increase in the reconstruction error with the b-value, as shown in Table~\ref{tab:bvalue_variants} for HCP.\footnote{An analogous table for ParkMR is omitted here for redundancy but is provided in the Supplemental material.} We hypothesise that this is due to two factors: (1) higher b-value shells are known to suffer more from noise; 
 (2) the acquired datasets contain different numbers of diffusion directions per shell, which might introduce an imbalance in optimisation during INR training.

\begin{table}[H]
  \centering
  \small
  \caption{Overall reconstruction performance (mean ± std, averaged across all voxels and shells) on HCP subjects. Training time for the pretrained setups is a factor of 1/4 compared to the not pretrained ones.}
  \label{tab:overall_variants}
  \begin{tabular}{lccccc}
    \toprule
    Training setup &  MSE ($\times 10^{-4}$) $  \downarrow$ & MAE $\downarrow$ & RE (\%) $\downarrow$& PSNR (dB) $\uparrow$ & SSIM $\uparrow$\\
    \midrule
Unregistered
  & $0.82\!\pm\!0.18$
  & $0.0065\!\pm\!0.00078$
  & $20.43\!\pm\!0.39$
  & $29.25\!\pm\!0.64$
  & $0.989\!\pm\!0.002$ \\
Pretrained\,Unreg.  & $0.92\!\pm\!0.21$
                     & $0.0069\!\pm\!0.00080$
                     & $21.56\!\pm\!0.31$
                     & $27.09\!\pm\!0.58$
                     & $0.844\!\pm\!0.009$ \\

Registered           & $0.40\!\pm\!0.05$
                     & $0.0047\!\pm\!0.00029$
                     & $5.30\!\pm\!1.15$
                     & $35.46\!\pm\!0.65$
                     & $0.989\!\pm\!0.001$ \\
Pretrained Reg.    & $0.70\!\pm\!0.07$
                     & $0.0059\!\pm\!0.00037$
                     & $6.10\!\pm\!1.26$
                     & $33.49\!\pm\!0.56$
                     & $0.986\!\pm\!0.001$ \\
    \bottomrule
  \end{tabular}
\end{table}

\begin{table}[H]
    \centering
    \small
    \caption{Overall reconstruction performance (mean ± std across subjects, averaged across all voxels and shells) for 26 subjects from the ParkMR Dataset.}\label{tab:overall_variants_ParkMR}
\begin{tabular}{lccccc}
    \toprule
    Training setup &  MSE ($\times 10^{-4}$) $\downarrow$ & MAE $\downarrow$ & RE (\%) $\downarrow$ & PSNR (dB) $\uparrow$ & SSIM $\uparrow$ \\
    \midrule
Unregistered
  & $5.80\!\pm\!0.15$
  & $0.0146\!\pm\!0.0018$
  & $15.88\!\pm\!3.92$
  & $26.39\!\pm\!1.17$
  & $0.972\!\pm\!0.005$ \\
Pretrained\,Unreg.
  & $10.42\!\pm\!1.73$
  & $0.0195\!\pm\!0.0018$
  & $22.22\!\pm\!6.16$
  & $24.13\!\pm\!1.46$
  & $0.950\!\pm\!0.008$ \\

Registered
  & $1.77\!\pm\!0.26$
  & $0.0092\!\pm\!0.0007$
  & $7.52\!\pm\!1.07$
  & $29.42\!\pm\!0.51$
  & $0.972\!\pm\!0.003$ \\

Pretrained Reg.
  & $2.37\!\pm\!0.33$
  & $0.0103\!\pm\!0.0008$
  & $7.95\!\pm\!1.23$
  & $27.97\!\pm\!0.59$
  & $0.963\!\pm\!0.004$ \\

    \bottomrule
\end{tabular}

\end{table}

\begin{table}[H]
  \centering
  \small
  \setlength{\tabcolsep}{3pt}  
  \caption{Reconstruction quality per b-value for the four training setups
           (mean ± std across subjects) for the HCP data. }
  \label{tab:bvalue_variants}
  \begin{tabular}{llccccc}
    \toprule
    \multirow{2}{*}{Training setup} & \multirow{2}{*}{b-value} &
    \multicolumn{5}{c}{Metric} \\ \cmidrule(lr){3-7}
& & MSE ($\times 10^{-4}$) $  \downarrow$ & MAE $\downarrow$ & RE (\%) $\downarrow$& PSNR (dB) $\uparrow$ & SSIM $\uparrow$ \\

    \midrule
\multirow{4}{*}{Unregistered}
    &   0   & $1.70\!\pm\!0.37$
              & $0.0095\!\pm\!0.0011$
              & $17.94\!\pm\!2.01$
              & $37.28\!\pm\!0.95$
              & $0.991\!\pm\!0.001$ \\
    & 1000  & $0.64\!\pm\!0.16$
              & $0.0062\!\pm\!0.00080$
              & $16.11\!\pm\!0.42$
              & $28.15\!\pm\!0.85$
              & $0.987\!\pm\!0.002$ \\
    & 2000  & $0.52\!\pm\!0.14$
              & $0.0056\!\pm\!0.00074$
              & $22.09\!\pm\!0.55$
              & $25.85\!\pm\!0.91$
              & $0.988\!\pm\!0.002$ \\
    & 3000  & $0.41\!\pm\!0.11$
              & $0.0049\!\pm\!0.00068$
              & $25.52\!\pm\!0.38$
              & $25.71\!\pm\!0.95$
              & $0.990\!\pm\!0.002$ \\
    \midrule
    \multirow{4}{*}{Pretrained\,Unreg.}
    &   0   & $1.89\!\pm\!0.42$
              & $0.0100\!\pm\!0.0011$
              & $20.74\!\pm\!2.38$
              & $26.30\!\pm\!0.96$
              & $0.919\!\pm\!0.012$ \\
    & 1000  & $0.75\!\pm\!0.18$
              & $0.0067\!\pm\!0.0008$
              & $17.01\!\pm\!0.46$
              & $27.17\!\pm\!1.35$
              & $0.842\!\pm\!0.012$ \\
    & 2000  & $0.58\!\pm\!0.15$
              & $0.0059\!\pm\!0.0008$
              & $22.59\!\pm\!0.67$
              & $27.11\!\pm\!0.91$
              & $0.808\!\pm\!0.016$ \\
    & 3000  & $0.45\!\pm\!0.13$
              & $0.0050\!\pm\!0.0007$
              & $25.91\!\pm\!0.48$
              & $27.77\!\pm\!1.07$
              & $0.806\!\pm\!0.010$ \\
    \midrule

\multirow{4}{*}{Registered}
    &   0   & $0.29\!\pm\!0.03$
              & $0.0040\!\pm\!0.0002$
              & $2.89\!\pm\!0.60$
              & $45.42\!\pm\!0.43$
              & $0.996\!\pm\!0.001$ \\
    & 1000  & $0.55\!\pm\!0.07$
              & $0.0055\!\pm\!0.0004$
              & $5.74\!\pm\!1.25$
              & $34.24\!\pm\!0.87$
              & $0.986\!\pm\!0.001$ \\
    & 2000  & $0.44\!\pm\!0.06$
              & $0.0049\!\pm\!0.0004$
              & $6.40\!\pm\!1.51$
              & $31.35\!\pm\!1.00$
              & $0.985\!\pm\!0.002$ \\
    & 3000  & $0.33\!\pm\!0.05$
              & $0.0042\!\pm\!0.0003$
              & $6.19\!\pm\!1.53$
              & $30.85\!\pm\!1.00$
              & $0.988\!\pm\!0.002$ \\
    \midrule
\multirow{4}{*}{Pretrained Reg.}
    &   0   & $1.31\!\pm\!0.12$
              & $0.0083\!\pm\!0.00039$
              & $5.17\!\pm\!0.91$
              & $38.83\!\pm\!0.39$
              & $0.988\!\pm\!0.001$ \\
    & 1000  & $0.68\!\pm\!0.09$
              & $0.0061\!\pm\!0.00042$
              & $6.31\!\pm\!1.35$
              & $33.28\!\pm\!0.83$
              & $0.983\!\pm\!0.001$ \\
    & 2000  & $0.47\!\pm\!0.07$
              & $0.0051\!\pm\!0.00041$
              & $6.62\!\pm\!1.54$
              & $31.07\!\pm\!0.91$
              & $0.984\!\pm\!0.002$ \\
    & 3000  & $0.34\!\pm\!0.06$
              & $0.0043\!\pm\!0.00039$
              & $6.31\!\pm\!1.56$
              & $30.75\!\pm\!0.90$
              & $0.988\!\pm\!0.002$ \\

    \bottomrule
  \end{tabular}
\end{table}

\subsection{Q-Space undersampling: Number of training points (b-vectors)}
\begin{figure}[]
    \centering
    \includegraphics[width=0.6\linewidth]{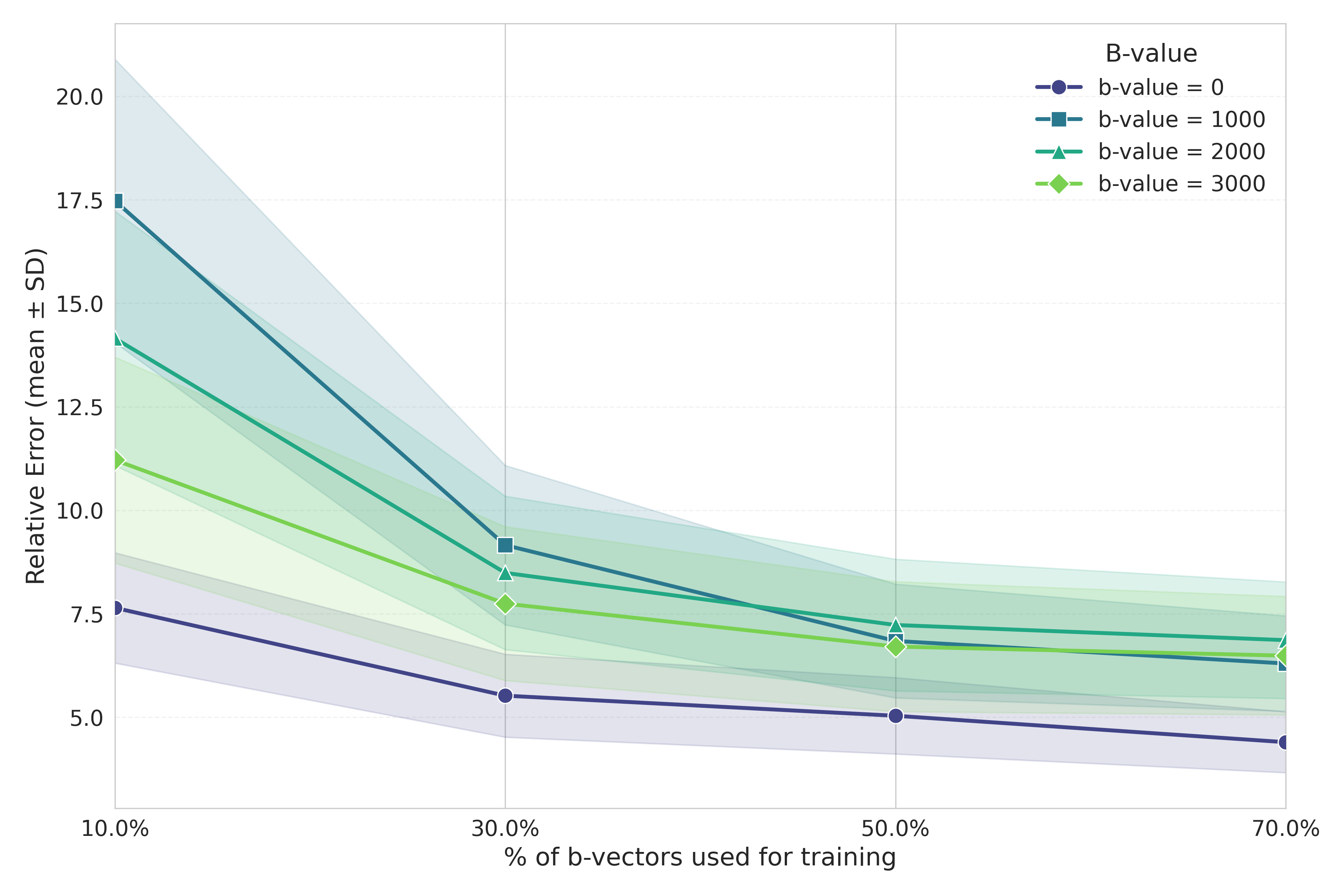}
  \caption{Relative error (mean $\pm$\,std) of the \textsc{Wire} model when trained with 10-70\% of diffusion directions for the registered HCP data with 288 gradient directions in total. }
    \label{fig:bvec_interp}
\end{figure}

To evaluate the robustness of the \textsc{Wire} model to sparse q-space sampling, we first investigate how reconstruction accuracy is affected by the number of available training b-vectors. 
We trained the model using stratified samples of 10\%, 30\%, 50\%, and 70\% of the total 288 unique b-value b-vector pairs from HCP, testing its predictive accuracy on the remaining 30\% of held-out directions. This analysis seeks to determine the data efficiency of our INR approach and the minimal q-space sampling density required for reliable signal representation. As shown in Fig. \ref{fig:bvec_interp} and Table \ref{tab:b_vector_interp}, error drops steeply from 10\% to 30\% of b-vectors and plateaus beyond 50\%; all shells fall below 10\% error at 30\%. Note that for the b = 0 shell, while the underlying data remains constant across different b-vectors, the varied training splits influence the model's exposure to b = 0 features, which is reflected in the reconstruction error.

\begin{table}
  \centering
  \small
  \caption{Overall reconstruction quality on the HCP subset for the \textsc{Wire} model with varying fractions of gradient directions used for training (mean\,±\,std across subjects)}
  \label{tab:b_vector_interp}
  \begin{tabular}{lccccc}
    \toprule
    \% of training & MSE ($\times 10^{-4}$) $\downarrow$ & MAE $\downarrow$ & RE (\%) $\downarrow$ & PSNR (dB) $\uparrow$ & SSIM $\uparrow$ \\
    \midrule
    10\% & 3.63 ± 0.70 & 0.0125 ± 0.0013 & 12.6 ± 2.6 & 26.7 ± 0.5 & 0.950 ± 0.006 \\
    30\% & 1.17 ± 0.13 & 0.0074 ± 0.0004 & 7.7 ± 1.6  & 31.4 ± 0.6 & 0.978 ± 0.001 \\
    50\% & 0.74 ± 0.14 & 0.0062 ± 0.0006 & 6.5 ± 1.4  & 33.1 ± 0.7 & 0.984 ± 0.002 \\
    70\% & 0.60 ± 0.10 & 0.0056 ± 0.0005 & 6.0 ± 1.2  & 33.8 ± 0.5 & 0.986 ± 0.002 \\
    \bottomrule
  \end{tabular}
\end{table}

\subsection{Shell interpolation}
In addition to evaluating performance with varying numbers of b-vector samples within individual shells, we assessed the \textsc{Wire} model's capability to interpolate the diffusion signal onto an unobserved shell, specifically $b=2000 \text{ s/mm}^2$. To mitigate overfitting to the training shells and encourage generalisation to this unseen b-value, we incorporated L2-regularisation and a dropout strategy for b-values during training (more details in the Supplementary material). For the target b = 2000 $\text{ s/mm}^2$ shell, the model achieved a MSE of $10.85 \pm 4.51 \times 10^{-4}$, a MAE of $0.0230 \pm 0.0053$, and a RE of $22.89 \pm 3.91 \%$. In terms of image quality metrics, this corresponded to a PSNR of $26.02 \pm 0.96 \text{ dB}$ and a SSIM of $0.894 \pm 0.028$. All metrics are reported for the HCP subset with 10 subjects. 

\begin{figure}
    \centering
    \includegraphics[width=0.65\linewidth]{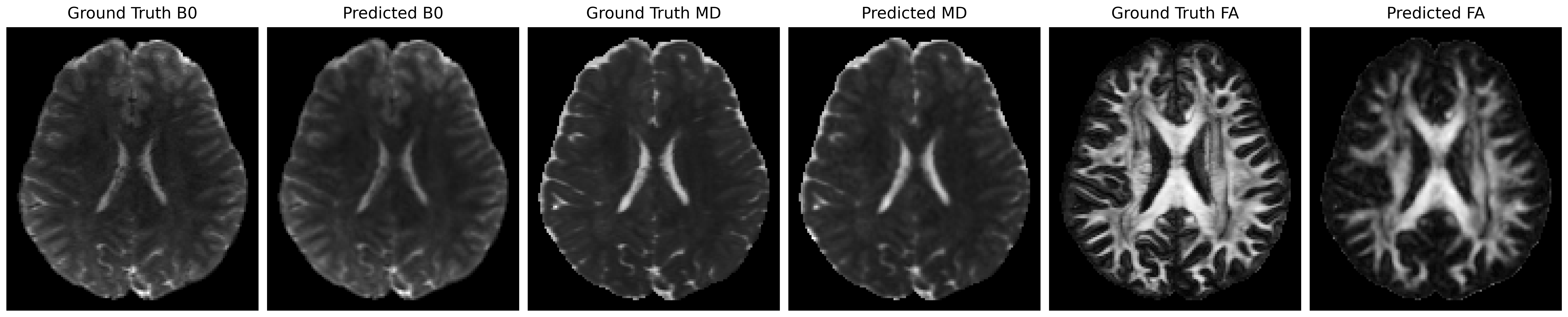}
    \caption{Visualisation of predicted and ground truth B0, MD and FA for a HCP subject.}
    \label{fig:vis_md_fa}
\end{figure}
\subsection{Downstream metrics: Mean diffusivity and fractional anisotropy}
Computing downstream representations of DWI data is common, and to evaluate the practical relevance and quality of the underlying signal representation, we derive the mean diffusivity (MD) and fractional anisotropy (FA) from our INR-derived diffusion signals. Evaluation metrics, computed within the brain mask, are reported in Table \ref{tab:md_fa_reconstruction} and visualised in Fig. \ref{fig:vis_md_fa}. The analysis of downstream metrics was performed with the \textsc{Wire} model on registered and unregistered data and when employing pretrained model parameters. We report the fractional anisotropy in the white matter using the threshold of FA$\geq$ 0.2 on a voxel-level, in line with previous literature \cite{ghosh_combined_2017,coll_assessment_2020}. For the relative errors of MD, we remove voxels with signal less than 0.001. We note that the relative errors of the registered subjects are skewed, with some voxels on the edge of the brain mask posting larger errors.

\begin{table}
  \centering
  \small
  \setlength{\tabcolsep}{3pt} 
  \caption{Reconstruction quality for MD and FA (WM) parameters for the four training variants (mean $\pm$ std for MSE, MAE, Mean RE; mean for percentile RE) for the HCP subset. }
  \label{tab:md_fa_reconstruction}
  \begin{tabular}{llcccc}
    \toprule
    \multirow{2}{*}{Training setup} & \multirow{2}{*}{Parameter} &
    \multicolumn{4}{c}{Metric} \\ \cmidrule(lr){3-6}
    & & MSE & MAE & Mean RE (\%) & (25th RE, 75th RE) \% \\
    \midrule
    \multirow{2}{*}{Unregistered}
    & MD & $0.000 \pm 0.0000$ & $0.000061 \pm 0.000005$ & $7.35 \pm 0.45$ & ($2.56, 10.31$) \\
    & FA (WM) & $0.027 \pm 0.0039$ & $0.1371 \pm 0.0110$ & $34.41 \pm 2.82$ & ($17.14, 50.25$)
    \\
    \midrule
    \multirow{2}{*}{Pretrained Unreg}
    & MD & $0.000 \pm 0.0000$ & $0.000069 \pm 0.000006$ & $8.50 \pm 0.50$ & ($2.92, 12.00$) \\
    & FA (WM) & $0.039 \pm 0.0054$ & $0.1678 \pm 0.0131$ & $41.74 \pm 3.23$ & ($23.63, 59.50$) \\
    \midrule
    \multirow{2}{*}{Registered}
    & MD & $0.000 \pm 0.0000$ & $0.000021 \pm 0.000006$ & $4.37 \pm 0.77$ & ($1.55, 6.07$) \\
    & FA (WM) & $0.037 \pm 0.0097$ & $0.1126 \pm 0.0179$ & $19.15 \pm 2.92$ & ($5.94, 26.54$) \\
    \midrule
    \multirow{2}{*}{Pretrained Reg}
    & MD & $0.000 \pm 0.0000$ & $0.000030 \pm 0.000009$ & $7.83 \pm 1.13$ & ($2.63, 11.11$) \\
    & FA (WM) & $0.044 \pm 0.0064$ & $0.1224 \pm 0.0096$ & $20.53 \pm 1.64$ & ($6.32, 28.26$) \\
    \bottomrule
  \end{tabular}
\end{table}

\FloatBarrier

\section{Discussion} 


\subsection{Pretraining INRs for DWI}
When evaluating the reconstruction errors from subjects in original and template space, we note that INRs trained on subject-specific images that were nonlinearly registered to template space showed lower representation errors across different experiments. We assume this to be a side effect of the SH decomposition and re-sampling step, effectively acting as a smoothing operation in q-space. Specifically, projecting the signal onto a band-limited SH basis suppresses higher-order angular frequencies and may reduce local signal variability. Though necessary for reorienting and interpolating signals onto a common gradient scheme, this step can bias the experiments in favour of the registered subjects by removing high-frequency q-space details that are harder to model.

\subsection{Limitations and future work}
Our current model has several limitations that we plan to address in future work. First, hyperparameters were selected empirically, making it important to assess their impact on INR performance and compression. We also plan to improve the framework as follows:

\paragraph{Intershell interpolation}
Overfitting to the data is an intrinsic property of INRs, yet the goal is to render a continuous representation in q-space. In the results, we note that the ability to interpolate b-vectors is better than that for b-values. This is likely due to the data being acquired at only a few discrete b-values. We mitigate this through small perturbations and regularisation, but obtaining a smooth representation over the b-values remains challenging. Constraining a smooth representation through a physical model could be a promising direction for INRs for DWI prediction.


\paragraph{Intrashell interpolation}
In the Supplemental material, we present a comparison with the traditional SHORE model \cite{ozarslan_simple_2009,ozarslan_mean_2013} for the task of b-vector interpolation. While our model performs adequately on this task, the specific modelling of the SHORE model is still able to achieve higher precision in the reconstructions. This suggests that general data-driven methods are not yet on the level of specific hand-engineered feature representations. In our work, we focus on improving training speed, but increasing the expressivity of the model is likely to bridge this gap.

\section{Conclusion}

We introduce a novel INR approach, alongside a group-averaged DWI template, for predicting the diffusion MRI signal. Our findings show that pretraining INRs on this template, particularly when subjects are registered to it, accelerates learning for new subjects. Furthermore, experiments show promise for q-space interpolation from sparse data, and deriving downstream metrics like mean diffusivity and fractional anisotropy from reconstructions. While INRs inherently involve a frequency tradeoff, this can be mitigated through architectural choices. 
Collectively, our results affirm INRs as a compelling direction for analysis 
of diffusion MRI data.

\newpage
\appendix
\section*{\LARGE Supplemental Material}
\section{Code availability}
We have made, training code and configurations available here: \url{https://github.com/anonymous4git/binr}
\section{Dataset details}\label{suppl_data}

\subsection{Preprocessing and Registration: HCP}

For the HCP dataset, we used the minimally preprocessed data from the Human Connectome Project (HCP), which includes correction for susceptibility distortions, motion, eddy currents, and alignment to each subject’s T1-weighted image \cite{glasser_minimal_2013}. To enable the transformation to MNI space, we reused the affine and nonlinear warp fields provided by the HCP pipeline. These transforms were converted to MRtrix3-compatible format using axis-specific reordering and adjusted for \texttt{element\_precision} compatibility.

Shell-specific amplitudes were converted to spherical harmonics (SH) using \texttt{amp2sh} with $l_\text{max} = 8$. The resulting SH-FODs were transformed using \texttt{mrtransform -reorient\_fod yes} with the converted warp and affine. Following transformation, all SH volumes were projected onto a predefined, shell-specific uniform gradient scheme. The mean $b_0$ image was duplicated to match the number of gradient directions per shell.

\subsection{Preprocessing and Registration: ParkMR}


The MD-dMRI images were acquired with a protocol that included a linear tensor encoding (LTE) and a spherical tensor encoding (STE); however, for the present work, we only use the LTE data. The LTE data consisted of 43 images with varying $b$-values ($b = 0$, 100, 700, 1400, 2000~$\unit{s}/\unit{mm}^{2}$) and is described in~\cite{SZCZEPANKIEWICZ2019104208}. Image parameters were: TR~$=4000$\,ms, echo time (TE)~$=111$\,ms, flip angle~$=90^{\circ}$, bandwidth~$=3311$\,Hz\,px$^{-1}$, and isotropic voxel size~$=2.5$\,mm. Topup corrections were applied using FSL’s \texttt{topup} tool~\cite{ANDERSSON2003870, ANDERSSON20161063} to compensate for EPI distortions, using a synthetic $b_0$-image created with Synb0-DISCO ~\cite{schilling2019synb0disco, schilling2020distortion}. For a more complete overview of MD-dMRI acquisition, see \cite{olsson_effects_2025}.

\begin{figure}[H]
    \centering
    \includegraphics[width=1.\linewidth]{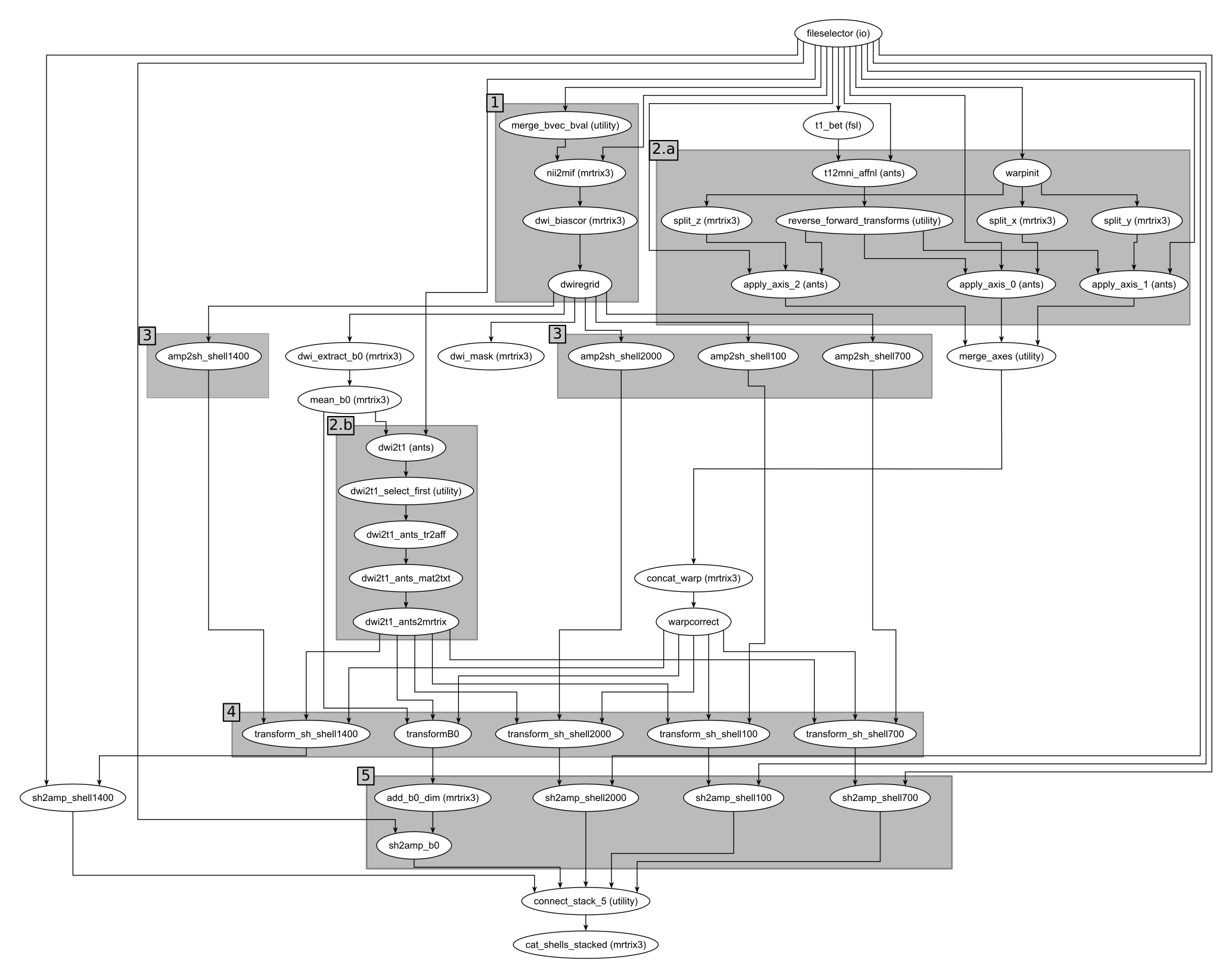}
    \caption{Signal flow diagram of the ParkMR nonlinear registration pipeline, exported from Nipype. The pipeline include the following stages: (1) general DWI preprocessing using MRtrix3; (2.a) nonlinear registration of T1 images to MNI 2\,mm space using ANTs with affine and SyN transformations; (2.b) rigid registration of the mean $b_0$ volume to the T1 image; (3) conversion of shell-specific DWI amplitudes to spherical harmonics (SH) with $l_\text{max}=8$; (4) spatial transformation of SH volumes using MRtrix3 with FOD reorientation; and (5) projection of SH signals onto a predefined, shell-specific gradient scheme, including duplication of the $b_0$ image. Nodes use MRtrix3, ANTs, and utility functions within Nipype.}
    \label{fig:reg-flow}
\end{figure}

Following the processing above, the data was used directly in non-registered experiments. For the multishell DWI registration, we ran the pipeline shown in Figure~\ref{fig:reg-flow} with the parameters listed below.

\paragraph{1. General DWI preprocessing.} Preprocessing includes gradient merging and conversion (MRtrix3), followed by denoising, Gibbs ringing removal, and FSL-based bias correction using \texttt{DWIBiasCorrect(use\_fsl=True)}. Images are regridded to 1.25\,mm isotropic resolution using \texttt{MRRegrid} with cubic interpolation.

\paragraph{2.a. T1 to MNI nonlinear registration.} Performed with \texttt{ants.Registration} (see main paper, Ref. [43,44]) using affine and SyN transforms. Parameters: 
\texttt{transform\_parameters = [(2.0,), (0.25, 3.0, 0.0)]}, 
\texttt{number\_of\_iterations = [[1500, 200], [100, 50, 30]]}, 
\texttt{metric = ['Mattes']}, 
\texttt{radius\_or\_number\_of\_bins = 32}, 
\texttt{smoothing\_sigmas = [[1, 0], [2, 1, 0]]}, 
\texttt{shrink\_factors = [[2, 1], [3, 2, 1]]}, 
\texttt{use\_histogram\_matching = True}, 
\texttt{collapse\_output\_transforms = True}. 
The transform is converted to MRtrix3 format.

\paragraph{2.b. $b_0$ to T1 registration.} The mean $b_0$ image is extracted and registered to T1 using \texttt{ants.Registration} (see main paper, Ref. [43,44]) with a rigid transform. Parameters: 
\texttt{transform = ['Rigid']}, 
\texttt{sampling\_strategy = ['Random']}, 
\texttt{sampling\_percentage = [0.05]}, 
\texttt{number\_of\_iterations = [[500, 100]]}. 
The resulting transform is converted to MRtrix3 format.

\paragraph{3. Spherical harmonic conversion.} Shell-specific DWI amplitudes are converted to SH using MRtrix3 \texttt{amp2sh} with $l_\text{max} = 8$. Shells: $b = 100$, $700$, $1400$, $2000$\,s/mm$^2$.

\paragraph{4. Transformation of SH data.} SH volumes are transformed to MNI space using \texttt{mrtransform -reorient\_fod yes}. Each shell is processed independently.

\paragraph{5. Resampling to common gradient scheme.} SH volumes are projected to a shell-specific uniform gradient scheme. The mean $b_0$ image is duplicated to match the number of directions per shell. Finaly, the channels representing different gradient directions and b-values are stacked into a 4D array and stored as NIFTII file.

\raggedbottom

\begin{table}[H]
\centering
\caption{Demographic and clinical characteristics of the study cohort for the MD-dMRI data. Values are reported as \emph{mean ± SD} with the range in square brackets. A dash (--) indicates “not applicable”.}
\label{tab:supp_demographics}
\begin{tabular}{lcc}
\toprule
\textbf{Subject Info} & \textbf{Controls} & \textbf{PD Patients} \\
\midrule
Mean age $\pm$ SD [min–max]        & $60 \pm 6\,[50\text{–}78]$   & $63 \pm 9\,[47\text{–}78]$ \\
Gender [F:M]                       & 3:14                         & 3:9                         \\
Disease duration [years]           & --                           & $6.4 \pm 5.3\,[1\text{–}17]$ \\
LEDD [mg]                          & --                           & $479 \pm 301\,[72\text{–}958]$ \\
H\,\&\,Y stage                      & --                           & $1.75 \pm 0.45\,[1\text{–}2]$ \\
MoCA                               & --                           & $28.9 \pm 1.2\,[27\text{–}30]$ \\
MDS–UPDRS–1                        & --                           & $7.0 \pm 2.3\,[2\text{–}10]$ \\
MDS–UPDRS–2                        & --                           & $6.5 \pm 2.3\,[3\text{–}11]$ \\
MDS–UPDRS–3                        & --                           & $18.7 \pm 11.0\,[4\text{–}41]$ \\
MDS–UPDRS–4                        & --                           & $5.1 \pm 4.1\,[0\text{–}12]$ \\
Total MDS–UPDRS                    & --                           & $35.3 \pm 9.8\,[24\text{–}58]$ \\
\bottomrule
\end{tabular}
\label{demo}
\end{table}

\FloatBarrier


\begin{table}
  \centering
  \small
  \setlength{\tabcolsep}{3pt}
  \caption{Quantitative metrics (mean ± std across subjects) for the ParkMR data. Lower is better for MSE/MAE/RE, higher is better for PSNR/SSIM.}
  \label{tab:bvalue_variants_park}
  \begin{tabular}{llccccc}
    \toprule
    \multirow{2}{*}{Training setup} & \multirow{2}{*}{b-value} &
    \multicolumn{5}{c}{Metric} \\ \cmidrule(lr){3-7}
    & & MSE ($\times 10^{-4}$) $  \downarrow$ & MAE $\downarrow$ & RE (\%) $\downarrow$& PSNR (dB) $\uparrow$ & SSIM $\uparrow$\\
    \midrule
\multirow{5}{*}{Unregistered}
    &   0   & $19.84\!\pm\!5.30$ & $0.0317\!\pm\!0.0041$ & $19.13\!\pm\!3.42$ & $27.04\!\pm\!1.15$ & $0.9546\!\pm\!0.0076$ \\
    & 100   & $5.55\!\pm\!1.52$  & $0.0164\!\pm\!0.0022$ & $11.69\!\pm\!2.33$ & $30.06\!\pm\!0.81$ & $0.9778\!\pm\!0.0035$ \\
    & 700   & $1.51\!\pm\!0.39$  & $0.0094\!\pm\!0.0012$ & $13.38\!\pm\!3.51$ & $25.47\!\pm\!1.89$ & $0.9733\!\pm\!0.0054$ \\
    & 1400  & $1.14\!\pm\!0.32$  & $0.0081\!\pm\!0.0012$ & $16.92\!\pm\!5.24$ & $24.58\!\pm\!1.98$ & $0.9761\!\pm\!0.0060$ \\
    & 2000  & $0.95\!\pm\!0.28$  & $0.0073\!\pm\!0.0011$ & $18.27\!\pm\!6.13$ & $24.78\!\pm\!2.11$ & $0.9786\!\pm\!0.0060$ \\
\midrule
\multirow{5}{*}{Registered}
    &   0   & $1.86\!\pm\!0.30$ & $0.0103\!\pm\!0.0008$ & $9.20\!\pm\!1.35$  & $28.15\!\pm\!1.12$ & $0.9620\!\pm\!0.0039$ \\
    & 100   & $4.28\!\pm\!0.67$ & $0.0145\!\pm\!0.0012$ & $8.55\!\pm\!1.02$  & $32.86\!\pm\!0.48$ & $0.9712\!\pm\!0.0027$ \\
    & 700   & $1.39\!\pm\!0.23$ & $0.0090\!\pm\!0.0008$ & $7.48\!\pm\!1.06$  & $28.27\!\pm\!0.77$ & $0.9670\!\pm\!0.0035$ \\
    & 1400  & $0.71\!\pm\!0.13$ & $0.0065\!\pm\!0.0006$ & $6.29\!\pm\!1.03$  & $28.50\!\pm\!0.73$ & $0.9786\!\pm\!0.0032$ \\
    & 2000  & $0.60\!\pm\!0.11$ & $0.0058\!\pm\!0.0006$ & $6.11\!\pm\!1.06$  & $29.32\!\pm\!0.73$ & $0.9808\!\pm\!0.0030$ \\
\midrule
\multirow{5}{*}{Pretrained + Reg.}
    &   0   & $2.53\!\pm\!0.38$ & $0.0111\!\pm\!0.0008$ & $9.34\!\pm\!1.55$  & $26.80\!\pm\!0.97$ & $0.9521\!\pm\!0.0047$ \\
    & 100   & $5.35\!\pm\!0.73$ & $0.0161\!\pm\!0.0011$ & $8.68\!\pm\!1.10$  & $31.87\!\pm\!0.46$ & $0.9625\!\pm\!0.0022$ \\
    & 700   & $2.05\!\pm\!0.35$ & $0.0104\!\pm\!0.0009$ & $8.14\!\pm\!1.23$  & $26.58\!\pm\!0.85$ & $0.9555\!\pm\!0.0050$ \\
    & 1400  & $1.04\!\pm\!0.21$ & $0.0074\!\pm\!0.0007$ & $6.98\!\pm\!1.23$  & $26.86\!\pm\!0.89$ & $0.9710\!\pm\!0.0046$ \\
    & 2000  & $0.86\!\pm\!0.17$ & $0.0066\!\pm\!0.0007$ & $6.64\!\pm\!1.25$  & $27.75\!\pm\!0.92$ & $0.9751\!\pm\!0.0043$ \\
\midrule
\multirow{5}{*}{Pretrained\,–\,Reg.}
    &   0   & $34.91\!\pm\!5.58$ & $0.0419\!\pm\!0.0037$ & $26.47\!\pm\!5.76$ & $24.50\!\pm\!0.80$ & $0.9109\!\pm\!0.0158$ \\
    & 100   & $11.57\!\pm\!2.13$ & $0.0245\!\pm\!0.0024$ & $19.16\!\pm\!4.54$ & $26.76\!\pm\!1.04$ & $0.9465\!\pm\!0.0087$ \\
    & 700   & $2.67\!\pm\!0.64$  & $0.0126\!\pm\!0.0015$ & $20.09\!\pm\!5.93$ & $22.96\!\pm\!2.16$ & $0.9553\!\pm\!0.0091$ \\
    & 1400  & $1.72\!\pm\!0.44$  & $0.0102\!\pm\!0.0013$ & $22.98\!\pm\!7.85$ & $22.76\!\pm\!2.19$ & $0.9643\!\pm\!0.0088$ \\
    & 2000  & $1.22\!\pm\!0.33$  & $0.0084\!\pm\!0.0012$ & $22.41\!\pm\!8.02$ & $23.64\!\pm\!2.18$ & $0.9723\!\pm\!0.0075$ \\
    \bottomrule
  \end{tabular}
\end{table}

\begin{table}
  \centering
  \small
  \setlength{\tabcolsep}{3pt}
  \caption{Reconstruction quality per b-value for the \textsc{Leaky},
           \textsc{Siren} and \textsc{Wire} models on the DWI template.
           Lower is better for MSE/MAE/RE, higher is better for PSNR/SSIM.}
  \label{tab:bvalue_metrics_template}
  \begin{tabular}{llccccc}
    \toprule
    \multirow{2}{*}{Method} & \multirow{2}{*}{b-value} &
      \multicolumn{5}{c}{Metric} \\ \cmidrule(lr){3-7}
    & & MSE $(\times10^{-4})$ $\downarrow$ & MAE $\downarrow$ & RE (\%) $\downarrow$ &
        PSNR (dB) $\uparrow$ & SSIM $\uparrow$\\
    \midrule
    \multirow{4}{*}{\textsc{Leaky}}
        &   0   & 2.24 & 0.0108 & 7.18 & 36.50 & 0.980 \\
        & 1000  & 4.07 & 0.0148 & 6.51 & 31.57 & 0.969 \\
        & 2000  & 2.22 & 0.0108 & 7.08 & 30.75 & 0.968 \\
        & 3000  & 1.58 & 0.0088 & 7.41 & 30.63 & 0.970 \\
    \midrule
    \multirow{4}{*}{\textsc{Siren}}
        & 0    & 1.09  & 0.0079 & 6.86 & 39.63 & 0.990 \\
        & 1000 & 2.56  & 0.0121 & 6.13 & 33.58 & 0.980 \\
        & 2000 & 2.06  & 0.0107 & 7.86 & 31.08 & 0.971 \\
        & 3000 & 1.55  & 0.0092 & 8.53 & 30.70 & 0.971 \\
    \midrule
    \multirow{4}{*}{\textsc{Wire}}
        & 0    & \textbf{0.39} & \textbf{0.0047} & \textbf{3.60} & \textbf{44.04} & \textbf{0.996} \\
        & 1000 & \textbf{1.03} & \textbf{0.0077} & \textbf{3.78} & \textbf{37.55} & \textbf{0.992} \\
        & 2000 & \textbf{0.84} & \textbf{0.0069} & \textbf{4.89} & \textbf{34.97} & \textbf{0.987} \\
        & 3000 & \textbf{0.66} & \textbf{0.0060} & \textbf{5.47} & \textbf{34.40} & \textbf{0.986} \\
    \bottomrule
  \end{tabular}
\end{table}

\FloatBarrier
\section{Results visualisation}
In Figures ~\ref{fig:nonwarped_plot} and ~\ref{fig:warped_plot}, the \textsc{Wire} predictions and the acquired diffusion‐weighted images (DWI) are visualised shell-by-shell.  
We note that the registration introduces blurring of the diffusion signal because of how we apply the nonlinear registration, i.e. by fitting spherical harmonic coefficients with limited degree (lmax=8) and subsequent sampling of the directional signal components. In Table \ref{tab:bvalue_metrics_template}, the reconstruction errors for the separate template are reported. Further, the corresponding results for the ParkMR data are shown in Table \ref{tab:bvalue_variants_park} for the unregistered/registered and pretraining settings.

\begin{figure}[H]
    \centering
    \includegraphics[width=0.7\linewidth]{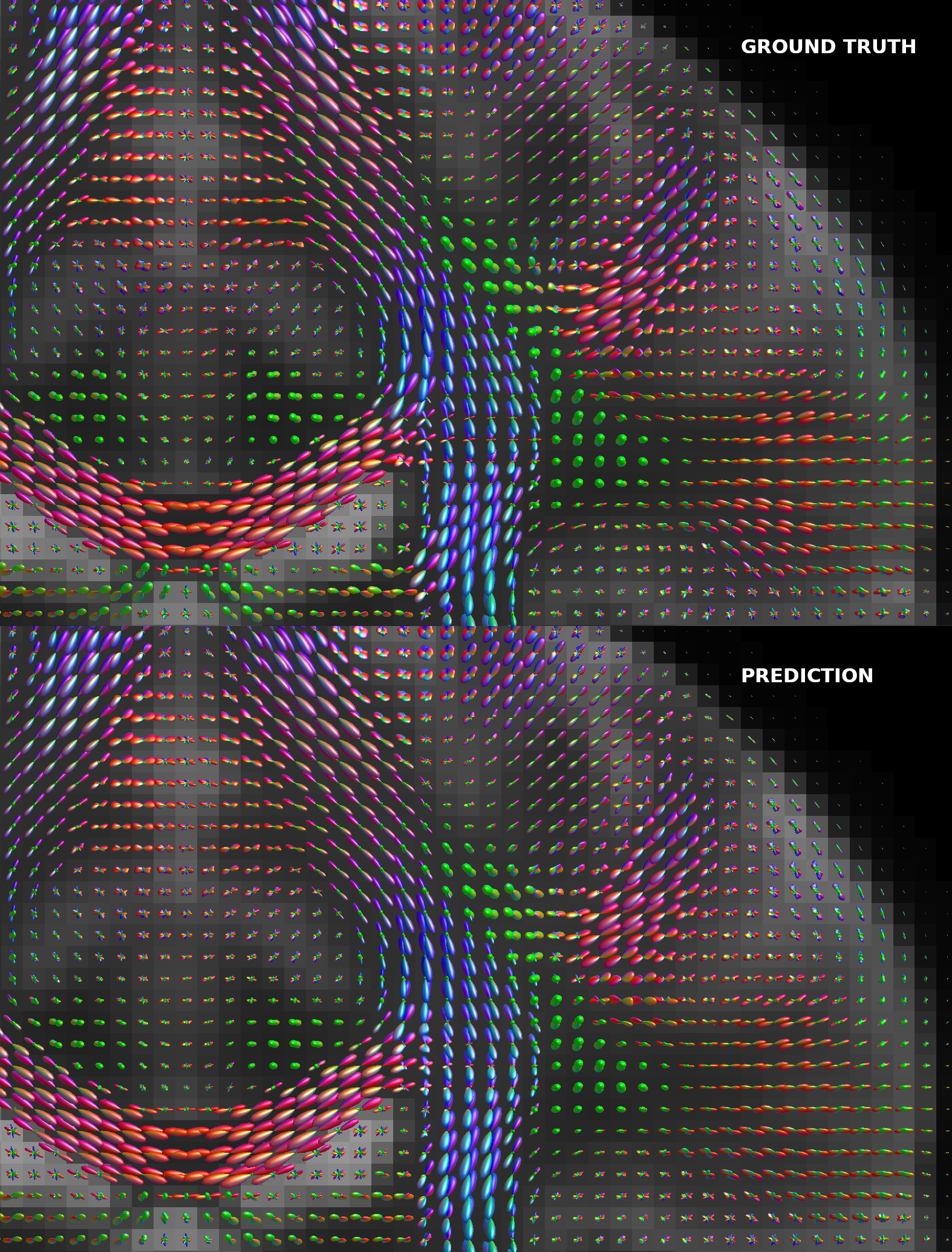}
    \caption{Fibre Orientation Distributions (FODs) from a coronal slice of the group-averaged DWI template. Ground truth FODs (top) and predictions from the \textsc{Wire} model (bottom) are shown.}
    \label{fig:fod}
\end{figure}

\begin{figure}
    \centering
    \includegraphics[width=0.8\linewidth]{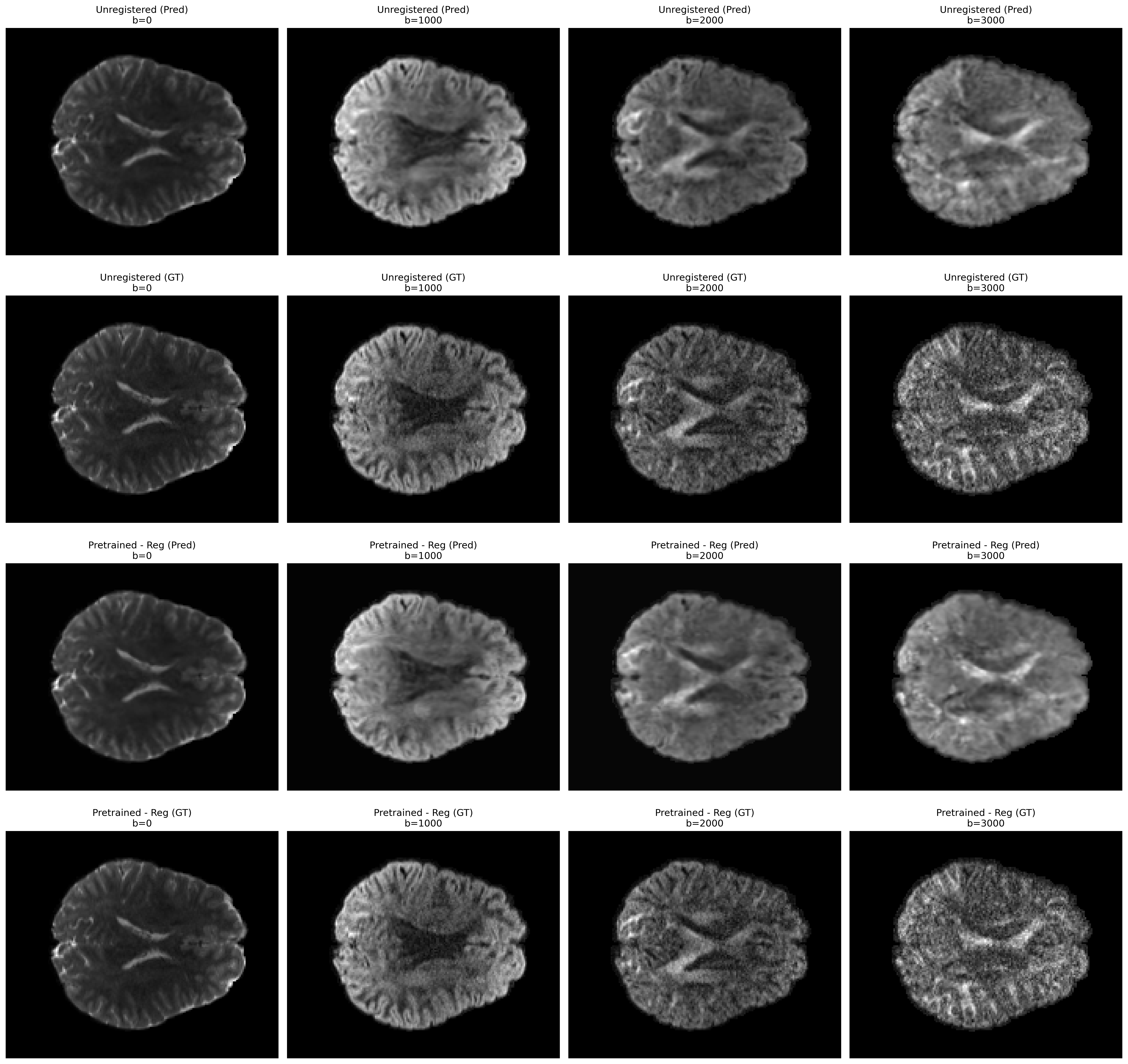}
    \caption{\textbf{Native space.}  
      Qualitative comparison in native DWI space (no template registration).  
      Four columns correspond to the target shells ($b=0$, $1000$, $2000$, $3000$ s/mm$^{2}$);  
      within each column, the top pair shows the INR prediction (\textit{Pred}) and the bottom pair the acquired ground truth (\textit{GT}).  
      Rows 1–2: INR trained from scratch in native space (\textit{Unregistered}).  
      Rows 3–4: template-initialised INR fitted in native space (\textit{Pretrained – Reg}).  
      Template pretraining reduces training time to one-quarter.}
    \label{fig:nonwarped_plot}
\end{figure}

\begin{figure}
    \centering
    \includegraphics[width=0.8\linewidth]{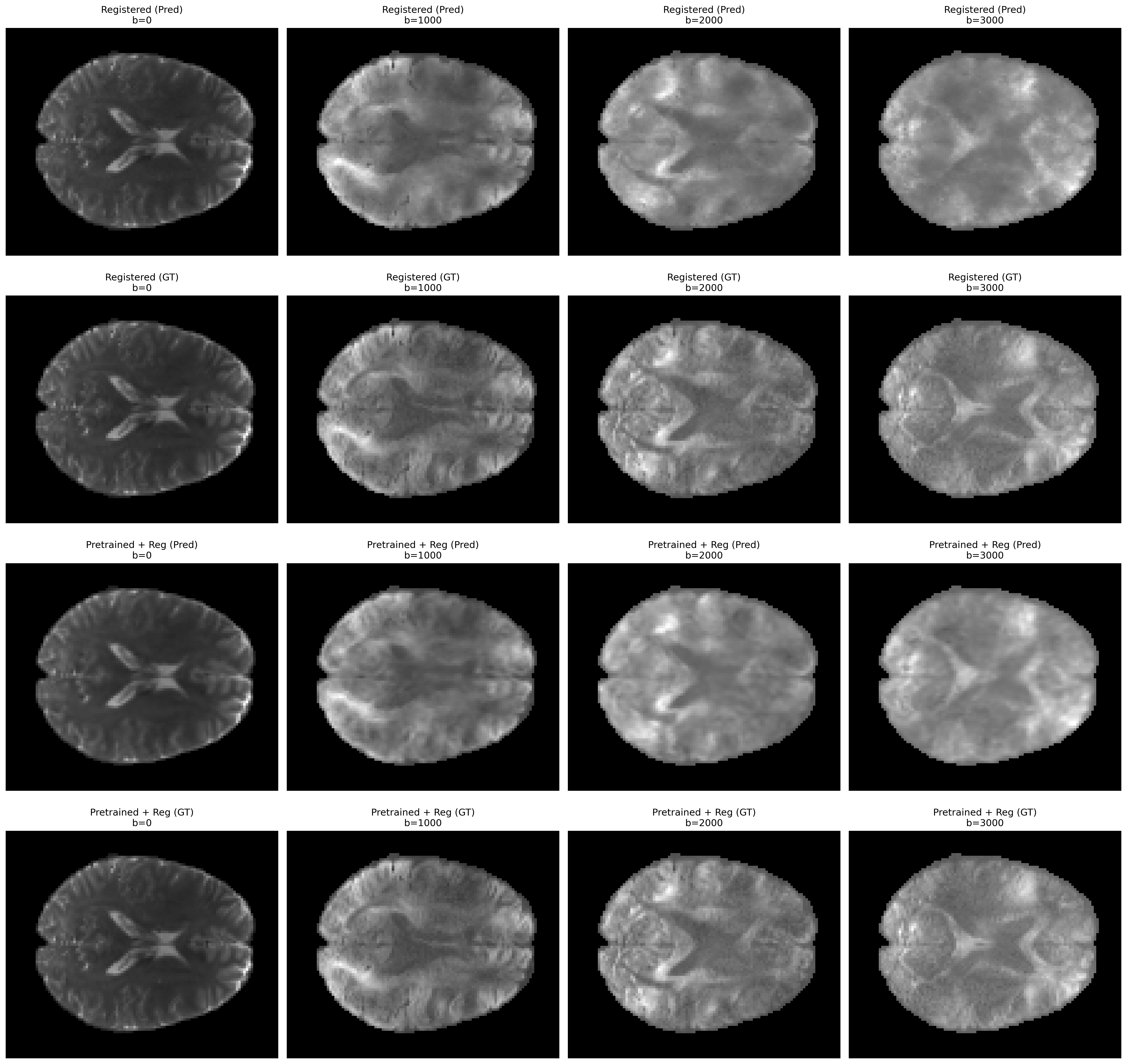}
    \caption{\textbf{Registered space.}  
      Same layout as Fig.~\ref{fig:nonwarped_plot}, but after rigid registration to the \mbox{$b\!=\!0$} volume.  
      Rows 1–2: INR trained from scratch in registered space (\textit{Registered}).  
      Rows 3–4: template-initialised INR (\textit{Pretrained + Reg}) with 1/4 of training time.}
    \label{fig:warped_plot}
\end{figure}

We also show a qualitative comparison of the FODs derived from the DWI predicted by the \textsc{Wire} model to ground truth FODs derived from the group-averaged DWI template in Figure \ref{fig:fod}. FODs were computed using multi-shell spherical deconvolution with the white matter response function provided by the authors of Ref. [29] in the main paper. Our DWI template averages data from 50 subjects, resulting in smoother signals than those typically observed in subject-specific FODs. 
\newpage

\raggedbottom
\section{Training configuration}
In this section, we detail the training configurations for all of our experiments. Our experiments were run on a compute cluster, but the suggested model can easily train quickly on a consumer-grade GPU such as an NVIDIA RTX 3090 using the same parameters as stated below. For re-training the model on an NVIDIA RTX 3090, the $418\,\text{k}$-parameter model ($\approx 1.7\,\text{MB}$ of weights) trains comfortably in memory, running $16$ epochs ($3\,424$ batches of $262\,\text{k}$ voxels) completed in about $6$ minutes, with roughly $23$ s per epoch on a single consumer-grade device.

\subsection{General training parameters}
In Table \ref{params_general}, the parameters for training that are equivalent for all experiments are described. Parameters specific to the \textsc{Wire} or \textsc{Siren} model or the feature encoding are listed in Table \ref{tab:model_params}. For the specific experiments we note any particular details below and for implementation-specific details we refer to our code repository.
\begin{table}[H]
\centering
\caption{Key hyperparameters for the model training and architecture. Variable names from the configuration are shown in \texttt{monospace}. The parameters that differ across the experiments are indicated by: \textit{varied}.}
\label{tab:hyperparameters_general} 
\begin{tabular}{lll}
\toprule
\textbf{Parameter Category} & \textbf{Hyperparameter} & \textbf{Value} \\
\midrule
\textit{Optimization} & Learning Rate (\texttt{lr}) & \textit{varied} \\
& Cosine LR Scheduler (\texttt{use\_cos\_lrsheduler}) & True \\
& Training Batch Size (\texttt{train\_batch\_size}) & 262,144 \\
& Epochs (\texttt{epochs}) & \textit{varied} \\
\midrule
\textit{Model Architecture} & Network Type (\texttt{network\_name}) & \textit{varied} \\
& Input Dimensionality (\texttt{in\_size}) & 301 \\
& Hidden Layer Dimensionality (\texttt{hidden\_size}) & 256 \\
& Output Dimensionality (\texttt{out\_size}) & 1 \\
& Number of Hidden Layers (\texttt{num\_layers}) & 3 \\
& Positional Encoder (\texttt{pos\_encoder}) & True \\
\midrule
\textit{Data Preprocessing} & Input Normalization (0-1) (\texttt{normalize\_0\_1}) & True \\
\bottomrule
\end{tabular}
\label{params_general} 
\end{table}

\begin{table}[h]
\centering
\begin{tabular}{lll}
\toprule
\textbf{Model Type} & \textbf{Parameter} & \textbf{Value} \\
\midrule
\textsc{Siren} & First Layer $\omega_0$ & 30.0 \\
              & Hidden Layer $\omega_h$ & 30.0 \\
\midrule
\textsc{Wire} & $\omega$ & 10.0 \\
               & $\sigma$ & 40.0 \\
\midrule
\textit{Shared Encoders} & Fourier Encoder Frequencies & 128 \\
                         & Fourier Feature Dim. & 256 \\
                         & Frequency Scaling Factor & 1.0 \\
                         & Spherical Harmonics $l_\text{max}$ & 8 (even bands only) \\
\bottomrule
\end{tabular}
\caption{Model-specific parameters for \textsc{Wire}, \textsc{Siren} models and feature encoding methods.}
\label{tab:model_params}
\end{table}

\subsection{DWI template training}
For the DWI template training comparisons, we train the \textsc{Leaky} ReLU model, \textsc{Siren} and \textsc{Wire} model for 64 epochs, with an initial learning rate of $0.001$.

\begin{table}[H]
\centering
\caption{Hyperparameters for training the INR on the group-averaged DWI template.}
\label{tab:hyperparameters_template}
\begin{tabular}{lll}
\toprule
\textbf{Parameter Category} & \textbf{Hyperparameter} & \textbf{Value} \\
\midrule
\textit{Optimization} & Learning Rate (\texttt{lr}) & $0.001$ \\
& Cosine LR Scheduler (\texttt{use\_cos\_lrsheduler}) & True \\
& Training Batch Size (\texttt{train\_batch\_size}) & 262,144 \\
& Epochs (\texttt{epochs}) & 64 \\
\midrule
\textit{Model Architecture} & Network Type (\texttt{network\_name}) & \textsc{Leaky} ReLU, \textsc{Siren}, \textsc{Wire} \\
& Input Dimensionality (\texttt{in\_size}) & 301 \\
& Hidden Layer Dimensionality (\texttt{hidden\_size}) & 256 \\
& Output Dimensionality (\texttt{out\_size}) & 1 \\
& Number of Hidden Layers (\texttt{num\_layers}) & 3 \\
& Positional Encoder (\texttt{pos\_encoder}) & True \\
\midrule
\textit{Data Preprocessing} & Input Normalization (0-1) (\texttt{normalize\_0\_1}) & True \\
\bottomrule
\end{tabular}
\label{params_template}
\end{table}

\subsection{Pretraining experiments}
In the experiments for comparing the effect of pretraining on the DWI template, the baseline is run for 64 epochs with both unregistered and registered subjects. The pretrained ones are trained for 16 epochs. All experiments use a learning rate of $0.0001$ and the \textsc{Wire} model.

\begin{table}[H]
\centering
\caption{Hyperparameters for pretraining and registration experiments.}
\label{tab:hyperparameters_pretraining}
\begin{tabular}{lll}
\toprule
\textbf{Parameter Category} & \textbf{Hyperparameter} & \textbf{Value} \\
\midrule
\textit{Optimization} & Learning Rate (\texttt{lr}) & $0.0001$ \\
& Cosine LR Scheduler (\texttt{use\_cos\_lrsheduler}) & True \\
& Training Batch Size (\texttt{train\_batch\_size}) & 262,144 \\
& Epochs (\texttt{epochs}) & Baseline: 64, Pretrained: 16 \\
\midrule
\textit{Model Architecture} & Network Type (\texttt{network\_name}) & \textsc{Wire} \\
& Input Dimensionality (\texttt{in\_size}) & 301 \\
& Hidden Layer Dimensionality (\texttt{hidden\_size}) & 256 \\
& Output Dimensionality (\texttt{out\_size}) & 1 \\
& Number of Hidden Layers (\texttt{num\_layers}) & 3 \\
& Positional Encoder (\texttt{pos\_encoder}) & True \\
\midrule
\textit{Data Preprocessing} & Input Normalization (0-1) (\texttt{normalize\_0\_1}) & True \\
\bottomrule
\end{tabular}
\label{params_pretraining}
\end{table}

\subsection{Interpolation of b-vectors}
In the experiment for interpolating b-vectors, we use the registered subjects and start training from the \textsc{Wire} DWI template checkpoint running 16 epochs in total with a learning rate of $0.001$.

\begin{table}[H]
\centering
\caption{Key hyperparameters for the b-vector interpolation experiments. }
\label{tab:hyperparameters_bvectors}
\begin{tabular}{lll}
\toprule
\textbf{Parameter Category} & \textbf{Hyperparameter} & \textbf{Value} \\
\midrule
\textit{Optimization} & Learning Rate (\texttt{lr}) & $0.001$ \\
& Cosine LR Scheduler (\texttt{use\_cos\_lrsheduler}) & True \\
& Training Batch Size (\texttt{train\_batch\_size}) & 262,144 \\
& Epochs (\texttt{epochs}) & 16 \\
\midrule
\textit{Model Architecture} & Network Type (\texttt{network\_name}) & \textsc{Wire} (from checkpoint) \\
& Input Dimensionality (\texttt{in\_size}) & 301 \\
& Hidden Layer Dimensionality (\texttt{hidden\_size}) & 256 \\
& Output Dimensionality (\texttt{out\_size}) & 1 \\
& Number of Hidden Layers (\texttt{num\_layers}) & 3 \\
& Positional Encoder (\texttt{pos\_encoder}) & True \\
\midrule
\textit{Data Preprocessing} & Input Normalization (0-1) (\texttt{normalize\_0\_1}) & True \\
\bottomrule
\end{tabular}
\label{params_bvectors}
\end{table}

\subsection{Interpolation of b-values}
For the b-value interpolation, we notice the model suffers from some overfitting to the sparse distribution of b-values. Therefore, we add three types of regularisation: dropout of b-values, perturbation of b-values and L2-regularisation. We use a dropout of 0.5 and uniformly randomly set the b-values to 0 during training. Further, we perturb the b-values during training by adding values from a distribution of $\mathcal{U}(-0.4b, 0.4b)$], where $b$ is the b-value. The strength of the regularisation is set to 0.0001. We run the interpolation experiments for 32 epochs.

\begin{table}[H]
\centering
\caption{Key hyperparameters for the b-value interpolation experiments. }
\label{tab:hyperparameters_bvalues}
\begin{tabular}{lll}
\toprule
\textbf{Parameter Category} & \textbf{Hyperparameter} & \textbf{Value} \\
\midrule
\textit{Optimization} & Learning Rate (\texttt{lr}) & 0.001 \\ 
& Cosine LR Scheduler (\texttt{use\_cos\_lrsheduler}) & True \\
& Training Batch Size (\texttt{train\_batch\_size}) & 262,144 \\
& Epochs (\texttt{epochs}) & 32 \\
& L2-regularisation strength & 0.0001 \\ 
& b-value dropout & 0.5 \\ 
& b-value perturbation & $\mathcal{U}(-0.4b, 0.4b)$] \\
\midrule
\textit{Model Architecture} & Network Type (\texttt{network\_name}) & \textsc{Wire} (from checkpoint) \\
& Input Dimensionality (\texttt{in\_size}) & 301 \\
& Hidden Layer Dimensionality (\texttt{hidden\_size}) & 256 \\
& Output Dimensionality (\texttt{out\_size}) & 1 \\
& Number of Hidden Layers (\texttt{num\_layers}) & 3 \\
& Positional Encoder (\texttt{pos\_encoder}) & True \\
\midrule
\textit{Data Preprocessing} & Input Normalization (0-1) (\texttt{normalize\_0\_1}) & True \\
\bottomrule
\end{tabular}
\label{params_bvalues}
\end{table}

\section{Comparison with SHORE model}

\begin{figure}[H]
    \centering
    \includegraphics[width=1.\linewidth]{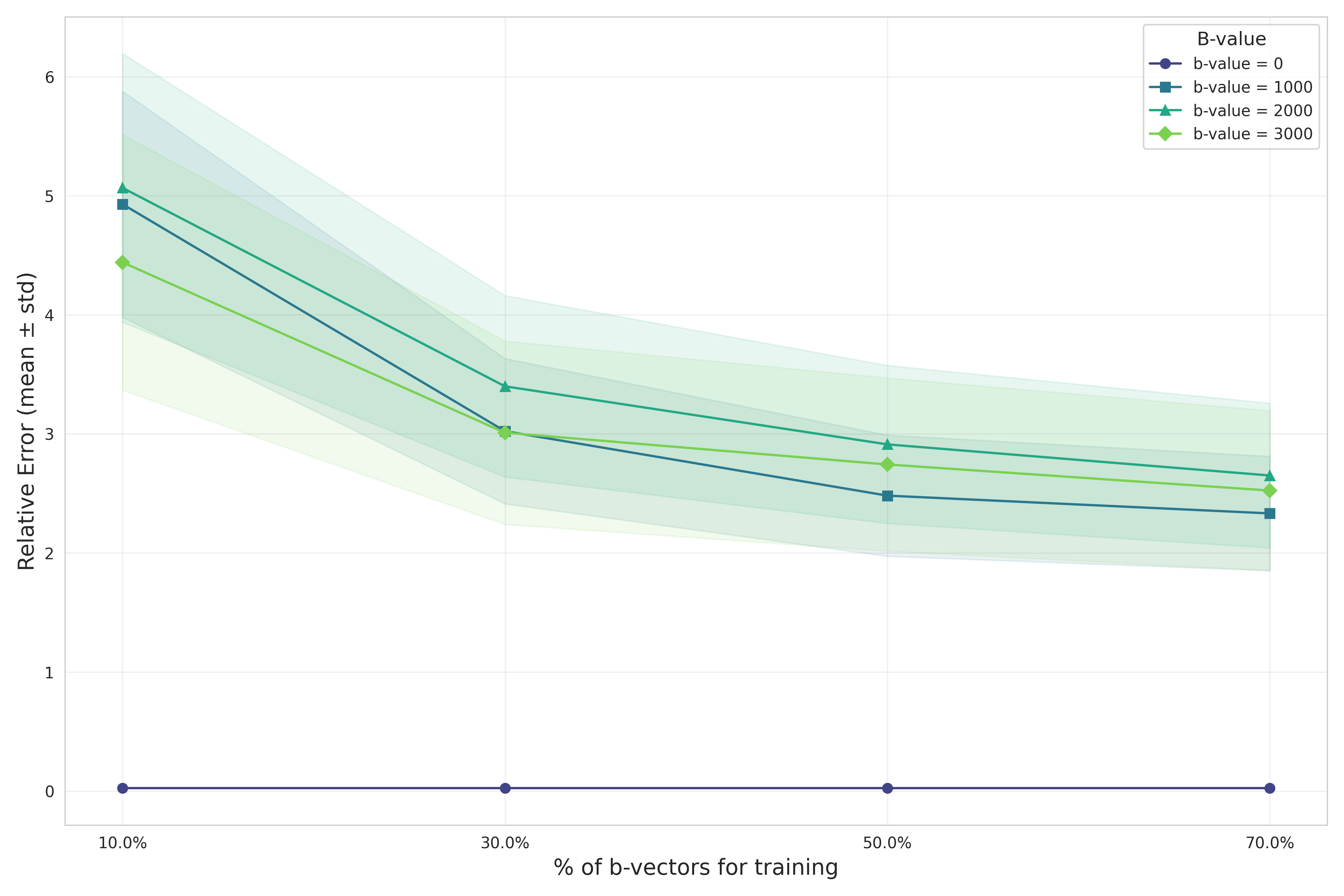}
    \caption{Relative error (mean ± std) of the SHORE model when fitted to 10–70\% of diffusion directions for the registered HCP data with 288 gradient directions in total. Data and setup correspond to the main paper, Figure 3.}
    \label{fig:re-shore}
\end{figure}

Figure \ref{fig:re-shore} shows the same experiment as in the main paper (Figure 3), but using a SHORE basis fitted independently for each voxel. While the relative error is lower than in our model, the comparison is not entirely direct: SHORE fits a separate set of coefficients per voxel, whereas our model learns a global representation across the entire volume. Note also that SHORE predicts diffusion signals relative to the b0 image, which is assumed to be known and thus incurs zero error. In future work, we aim to incorporate similar normalisation strategies to further improve performance.

\section{Superresolution experiment}

To analyze the generalization of our INR approach to the superresolution task, we trained the \textsc{Wire} model on the HCP subset downsampling training images 2x and 4x, respectively, and evaluated their performance on the original resolution. The hyperparameters used for these experiments are displayed in Table \ref{params_superres}.

\begin{table}[H]
\centering
\caption{Hyperparameters used for superresolution experiments.}
\label{tab:hyperparameters_bvectors}
\begin{tabular}{lll}
\toprule
\textbf{Parameter Category} & \textbf{Hyperparameter} & \textbf{Value} \\
\midrule
\textit{Optimization} & Learning Rate (\texttt{lr}) & $0.001$ \\
& Cosine LR Scheduler (\texttt{use\_cos\_lrsheduler}) & True \\
& Training Batch Size (\texttt{train\_batch\_size}) & 262,144 \\
& Epochs (\texttt{epochs}) & 32 \\
\midrule
\textit{Model Architecture} & Network Type (\texttt{network\_name}) & \textsc{Wire} (from checkpoint) \\
& Input Dimensionality (\texttt{in\_size}) & 301 \\
& Hidden Layer Dimensionality (\texttt{hidden\_size}) & 256 \\
& Output Dimensionality (\texttt{out\_size}) & 1 \\
& Number of Hidden Layers (\texttt{num\_layers}) & 3 \\
& Positional Encoder (\texttt{pos\_encoder}) & True \\
\midrule
\textit{Data Preprocessing} & Input Normalization (0-1) (\texttt{normalize\_0\_1}) & True \\
\bottomrule
\end{tabular}
\label{params_superres}
\end{table}

\begin{table}[H]
  \centering
  \small
  \setlength{\tabcolsep}{3pt}  
  \caption{Reconstruction quality per b-value (mean ± std across subjects) for 
           4× and 2× superresolution on the HCP data.}
  \label{tab:bvalue_sr}
  \begin{tabular}{llccccc}
    \toprule
    \multirow{2}{*}{Upscaling factor} & \multirow{2}{*}{b-value} &
    \multicolumn{5}{c}{Metric} \\ \cmidrule(lr){3-7}
     & & MSE ($\times 10^{-3}$) $\downarrow$ & MAE $\downarrow$ & RE (\%) $\downarrow$ & PSNR (dB) $\uparrow$ & SSIM $\uparrow$ \\
    \midrule
\multirow{4}{*}{4×}
     &   0   & $6.67\!\pm\!1.57$ & $0.051\!\pm\!0.010$ & $28.8\!\pm\!4.0$  & $21.9\!\pm\!1.0$ & $0.794\!\pm\!0.019$ \\
     & 1000  & $1.30\!\pm\!0.68$ & $0.027\!\pm\!0.008$ & $28.3\!\pm\!6.0$  & $20.8\!\pm\!1.4$ & $0.858\!\pm\!0.028$ \\
     & 2000  & $0.829\!\pm\!0.772$& $0.023\!\pm\!0.011$& $28.9\!\pm\!10.0$ & $19.6\!\pm\!2.5$ & $0.896\!\pm\!0.038$ \\
     & 3000  & $0.731\!\pm\!0.823$& $0.022\!\pm\!0.012$& $30.5\!\pm\!12.3$ & $18.9\!\pm\!3.2$ & $0.911\!\pm\!0.046$ \\
    \midrule
\multirow{4}{*}{2×}
     &   0   & $4.17\!\pm\!1.17$ & $0.039\!\pm\!0.010$ & $20.4\!\pm\!3.7$  & $24.0\!\pm\!1.2$ & $0.879\!\pm\!0.009$ \\
     & 1000  & $0.613\!\pm\!0.383$& $0.019\!\pm\!0.007$& $19.2\!\pm\!4.8$  & $24.3\!\pm\!1.7$ & $0.930\!\pm\!0.014$ \\
     & 2000  & $0.380\!\pm\!0.370$& $0.015\!\pm\!0.008$& $19.6\!\pm\!6.8$  & $23.0\!\pm\!2.4$ & $0.950\!\pm\!0.018$ \\
     & 3000  & $0.320\!\pm\!0.381$& $0.014\!\pm\!0.008$& $20.3\!\pm\!8.3$  & $22.3\!\pm\!2.9$ & $0.959\!\pm\!0.020$ \\
    \midrule
\multirow{4}{*}{No upscaling}
    &   0   & $0.29\!\pm\!0.03$
              & $0.0040\!\pm\!0.0002$
              & $2.89\!\pm\!0.60$
              & $45.42\!\pm\!0.43$
              & $0.996\!\pm\!0.001$ \\
    & 1000  & $0.55\!\pm\!0.07$
              & $0.0055\!\pm\!0.0004$
              & $5.74\!\pm\!1.25$
              & $34.24\!\pm\!0.87$
              & $0.986\!\pm\!0.001$ \\
    & 2000  & $0.44\!\pm\!0.06$
              & $0.0049\!\pm\!0.0004$
              & $6.40\!\pm\!1.51$
              & $31.35\!\pm\!1.00$
              & $0.985\!\pm\!0.002$ \\
    & 3000  & $0.33\!\pm\!0.05$
              & $0.0042\!\pm\!0.0003$
              & $6.19\!\pm\!1.53$
              & $30.85\!\pm\!1.00$
              & $0.988\!\pm\!0.002$ \\
    \bottomrule
  \end{tabular}
  \label{bval_superres}
\end{table}

\begin{table}[H]
  \centering
  \small
  \caption{Overall reconstruction performance (mean ± std, averaged across all voxels and shells) for the two upscaling factors.}
  \label{tab:overall_sr}
  \begin{tabular}{lccccc}
    \toprule
    Upscaling factor & MSE ($\times 10^{-3}$) $\downarrow$ & MAE $\downarrow$ & RE (\%) $\downarrow$ & PSNR (dB) $\uparrow$ & SSIM $\uparrow$ \\
    \midrule
      4× & $2.383\!\pm\!0.785$ & $0.031\!\pm\!0.009$ & $29.1\!\pm\!6.9$ & $20.3\!\pm\!1.7$ & $0.865\!\pm\!0.031$ \\
      2× & $1.372\!\pm\!0.472$ & $0.022\!\pm\!0.007$ & $19.9\!\pm\!5.1$ & $23.4\!\pm\!1.7$ & $0.930\!\pm\!0.014$ \\
No upscaling           & $0.40\!\pm\!0.05$
                     & $0.0047\!\pm\!0.00029$
                     & $5.30\!\pm\!1.15$
                     & $35.46\!\pm\!0.65$
                     & $0.989\!\pm\!0.001$ \\
    \bottomrule
  \end{tabular}
  \label{overall_superres}
\end{table}


The reconstruction error for the superresolution task per b-values is shown in Table \ref{bval_superres} and averaged over all shells in Table \ref{overall_superres}. We note that the realistic scenario of a 2x upscaling of the data, e.g., from a common clinical resolution of 2.5 mm to 1.25 mm, would maintain close structural similarity using the INR, especially for the higher b-values. More specialized upscaling techniques may be required for even greater upscaling.

\newpage
\bibliography{references,references_supplemental}

\end{document}